\documentclass[10pt]{revtex4}
\usepackage{graphicx}

\newcount\driver
\newcount\bozza

\font\ottorm=cmr8\font\ottoi=cmmi8\font\ottosy=cmsy8%
\font\ottocss=cmcsc8%
\font\sixrm=cmr6\font\sixi=cmmi6\font\sixsy=cmsy6%
\font\fiverm=cmr5\font\fivesy=cmsy5
\font\fivei=cmmi5
\font\tenmib=cmmib10
\font\sevenmib=cmmib10 scaled 800

 2

\font\cs=cmcsc10
\font\sc=cmcsc10

\font\ss=cmss10

\font\elevenrm=cmr11
\font\twelverm=cmr12
\font\ottorm=cmr8
\font\msytw=msbm9 scaled\magstep1

\font\msytwww=msbm5 scaled\magstep1
\font\indbf=cmbx10 scaled\magstep2

\font\ottorm=cmr8\font\ottoi=cmmi8\font\ottosy=cmsy8%
\font\ottocss=cmcsc8%
\font\sixrm=cmr6\font\sixi=cmmi6\font\sixsy=cmsy6%
\font\fiverm=cmr5\font\fivesy=cmsy5
\font\fivei=cmmi5

\def\ottopunti{\def\rm{\fam0\ottorm}%
\textfont0=\ottorm\scriptfont0=\sixrm\scriptscriptfont0=\fiverm%
\textfont1=\ottoi\scriptfont1=\sixi\scriptscriptfont1=\fivei%
\textfont2=\ottosy\scriptfont2=\sixsy\scriptscriptfont2=\fivesy%
\textfont4=\ottocss\scriptfont4=\sc\scriptscriptfont4=\sc%
\scriptfont4=\ottocss\scriptscriptfont4=\ottocss%
\textfont5=\tenmib\scriptfont5=\sevenmib\scriptscriptfont5=\fivei
\setbox\strutbox=\hbox{\vrule height7pt depth2pt width0pt}%
\normalbaselineskip=9pt\let\sc=\sixrm\normalbaselines\rm}

\mathchardef\BDpr = "0540  %Dpr
\mathchardef\Bg   = "050D  %gamma

{\count255=\time\divide\count255 by 60 \xdef\hourmin{\number\count255}
          \multiply\count255 by-60\advance\count255 by\time
     \xdef\hourmin{\hourmin:\ifnum\count255<10 0\fi\the\count255}}

\def\openone{\leavevmode\hbox{\elevenrm 1\kern-3.63pt\twelverm1}}%
\def\*{\vglue0.5truecm}

\let\a=\alpha   \let\g=\gamma  \let\d=\delta \let\e=\varepsilon
     \let\th=\theta \let\k=\kappa \let\l=\lambda
             \let\p=\pi    \let\r=\rho
\let\s=\sigma     
   \let\o=\omega
\let\G=\Gamma \let\D=\Delta  \let\L=\Lambda 
         
\let\O=\Omega 

\def\\{\hfill\break} \let\==\equiv

\let\io=\infty 

\let\0=\noindent

\def\ie{{\it i.e.}}\def\eg{{\it e.g.}}
\let\dpr=\partial

\def\tende#1{\,\vtop{\ialign{##\crcr\rightarrowfill\crcr
   \noalign{\kern-1pt\nointerlineskip} \hskip3.pt${\scriptstyle
   #1}$\hskip3.pt\crcr}}\,}
\def\circage{\lower2pt\hbox{$\,\buildrel > \over {\scriptstyle \sim}\,$}}
\def\otto{\,{\kern-1.truept\leftarrow\kern-5.truept\to\kern-1.truept}\,}
\def\fra#1#2{{#1\over#2}}

 \def\VV{{\cal V}}

\def\RR{{\cal R}}  
\def\DD{{\cal D}} \def\SS{{\cal S}}

\def\T#1{{#1_{\kern-3pt\lower7pt\hbox{$\widetilde{}$}}\kern3pt}}
\def\VVV#1{{\VV #1}_{\kern-3pt
\lower7pt\hbox{$\widetilde{}$}}\kern3pt\,}
\def\W#1{#1_{\kern-3pt\lower7.5pt\hbox{$\widetilde{}$}}\kern2pt\,}

\def\indica{\leaders \hbox to 0.5cm{\hss.\hss}\hfill}
\def\guida{\leaders\hbox to 1em{\hss.\hss}\hfill}

   \def\qq{{\bf q}}
   \def\pp{{\bf p}}
 \def\xx{{\bf x}} \def\yy{{\bf y}} 
\def\hhh{{\bf h}}
\def\kk{{\bf k}}
\def\mm{{\bf m}}\def\Vn{{\bf n}}

\def\VV#1{{\,\underline#1\,}}
\def\ul{\underline}

\def\defin{{\buildrel def\over=}}

\mathchardef\aa   = "050B
\mathchardef\bb   = "050C
\mathchardef\ggg  = "050D
\mathchardef\xxx  = "0518
\mathchardef\zzzzz= "0510
\mathchardef\oo   = "0521
\mathchardef\lll  = "0515
\mathchardef\Dp   = "0540
\mathchardef\H    = "0548
\mathchardef\FFF  = "0546
\mathchardef\ppp  = "0570
\mathchardef\Bn   = "0517
\mathchardef\pps  = "0520
\mathchardef\fff  = "0527
\mathchardef\FFF  = "0508
\mathchardef\nnnnn= "056E

\def\to{\rightarrow}

\def\qed{\hfill\raise1pt\hbox{\vrule height5pt width5pt depth0pt}}

\def\indic{\hbox{\raise-2pt \hbox{\indbf 1}}}

\def\rrr{\hbox{\msytwww R}}

 \def\ZZZ{\hbox{\msytw Z}}
 \def\zzz{\hbox{\msytwww Z}}

\def\ul#1{{\underline#1}}

\def\V0{{\bf 0}}

\font\tenmib=cmmib10 
\font\sevenmib=cmmib7\font\fivemib=cmmib5

\font\fivei=cmmi5\font\sixi=cmmi6\font\ottoi=cmmi8
\font\ottorm=cmr8\font\fiverm=cmr5\font\sixrm=cmr6
\font\ottosy=cmsy8\font\sixsy=cmsy6\font\fivesy=cmsy5%%
\font\ottocss=cmcsc8%

\mathchardef\Ba   = "050B  %alfa
\mathchardef\Bb   = "050C  %beta
\mathchardef\Bg   = "050D  %gamma
\mathchardef\Bd   = "050E  %delta
\mathchardef\Be   = "0522  %varepsilon
\mathchardef\Bee  = "050F  %epsilon
\mathchardef\Bz   = "0510  %zeta
\mathchardef\Bh   = "0511  %eta
\mathchardef\Bthh = "0512  %teta
\mathchardef\Bth  = "0523  %varteta
\mathchardef\Bi   = "0513  %iota
\mathchardef\Bk   = "0514  %kappa
\mathchardef\Bl   = "0515  %lambda
\mathchardef\Bm   = "0516  %mu
\mathchardef\Bn   = "0517  %nu
\mathchardef\Bx   = "0518  %xi
\mathchardef\Bom  = "0530  %omi
\mathchardef\Bp   = "0519  %pi
\mathchardef\Br   = "0525  %ro
\mathchardef\Bro  = "051A  %varrho
\mathchardef\Bs   = "051B  %sigma
\mathchardef\Bsi  = "0526  %varsigma
\mathchardef\Bt   = "051C  %tau
\mathchardef\Bu   = "051D  %upsilon
\mathchardef\Bf   = "0527  %phi
\mathchardef\Bff  = "051E  %varphi
\mathchardef\Bch  = "051F  %chi
\mathchardef\Bps  = "0520  %psi
\mathchardef\Bo   = "0521  %omega
\mathchardef\Bome = "0524  %varomega
\mathchardef\BG   = "0500  %Gamma
\mathchardef\BD   = "0501  %Delta
\mathchardef\BTh  = "0502  %Theta
\mathchardef\BL   = "0503  %Lambda
\mathchardef\BX   = "0504  %Xi
\mathchardef\BP   = "0505  %Pi
\mathchardef\BS   = "0506  %Sigma
\mathchardef\BU   = "0507  %Upsilon
\mathchardef\BF   = "0508  %Fi
\mathchardef\BPs  = "0509  %Psi
\mathchardef\BO   = "050A  %Omega
\mathchardef\BDpr = "0540  %Dpr
\mathchardef\Bstl = "053F  %*

\def\V#1{{\bf#1}}
\let\aa=\Ba\let\fff=\Bf\let\defin=\defi
\let\oo=\Bo\let\nn=\Bn
\let\pps=\Bps\def\hhh={\V h}
\let\bb=\Bb\def\ss{\ul{\s}}\def\uA{\ul{A}}

\def\rrr{\hbox{\msytwww R}}

 \def\ZZZ{\hbox{\msytw Z}}
 \def\zzz{\hbox{\msytwww Z}}

\let\ul=\underline

\def\const{{\rm const}}

\def\ins#1#2#3{\vbox to0pt{\kern-#2 \hbox{\kern#1 #3}\vss}\nointerlineskip}
\newdimen\xshift \newdimen\xwidth \newdimen\yshift
\newcount\griglia

\def\insertplot#1#2#3#4#5#6{%
\begin{figure}[h]
\begin{center}
\vspace{#2pt}
\begin{minipage}{#1pt}
#3
\ifnum\driver=1
\griglia=#6
\ifnum\griglia=1
\openout13=griglia.ps
\write13{gsave .2 setlinewidth}
\write13{0 10 #1 {dup 0 moveto #2 lineto } for}
\write13{0 10 #2 {dup 0 exch moveto #1 exch lineto } for}
\write13{stroke}
\write13{.5 setlinewidth}
\write13{0 50 #1 {dup 0 moveto #2 lineto } for}
\write13{0 50 #2 {dup 0 exch moveto #1 exch lineto } for}
\write13{stroke grestore}
\closeout13
\includegraphics{griglia.ps}\fi
\includegraphics{#4.ps}\fi
\ifnum\driver=2
\fi
\end{minipage}
\end{center}
\caption{#5}
\end{figure}
}

\newdimen\shift \shift=-1truecm
\def\lb#1{%
\ifnum\bozza=1
\label{#1}\rlap{\kern\shift{$\scriptstyle#1$}}
\else\label{#1}
\fi}

\def\be{\begin{equation}}
\def\ee{\end{equation}}
\def\bea{\begin{eqnarray}}\def\eea{\end{eqnarray}}
\def\bean{\begin{eqnarray*}}\def\eean{\end{eqnarray*}}
\def\bfr{\begin{flushright}}\def\efr{\end{flushright}}
\def\bc{\begin{center}}\def\ec{\end{center}}
\def\ba#1{\begin{array}{#1}} \def\ea{\end{array}}
\def\bd{\begin{description}}\def\ed{\end{description}}
\def\bv{\begin{verbatim}}\def\ev{\end{verbatim}}
\def\nn{\nonumber}
\def\Halmos{\hfill\vrule height10pt width4pt depth2pt \par\hbox to \hsize{}}

\renewcommand{\theequation}{\arabic{section}.\arabic{equation}}

\newdimen\xshift \newdimen\xwidth \newdimen\yshift \newdimen\ywidth

\def\ins#1#2#3{\vbox to0pt{\kern-#2\hbox{\kern#1 #3}\vss}\nointerlineskip}

\def\eqfig#1#2#3#4#5{
\par\xwidth=#1 \xshift=\hsize \advance\xshift
by-\xwidth \divide\xshift by 2
\yshift=#2 \divide\yshift by 2
\line{\hglue\xshift \vbox to #2{\vfil
#3 \includegraphics{#4.ps}
}\hfill\raise\yshift\hbox{#5}}}

\def\8{\write12}

\begin{document}

\title{Striped phases in two dimensional dipole systems}

\author{Alessandro Giuliani}\thanks{On leave from Dipartimento di Matematica
di Roma Tre, Largo S. Leonardo Murialdo 1, 00146 Roma, Italy.}
\affiliation{Department of Physics,
Princeton University, Princeton 08544 NJ, USA}
\author{Joel L. Lebowitz}
\affiliation{Department of Mathematics and Physics, Rutgers University,
Piscataway, NJ 08854 USA.}
\author{Elliott H. Lieb}
\affiliation{Departments of Mathematics and Physics,
Princeton University, Princeton, NJ 08544 USA.}
\vspace{1cm}
\date{\today}
\begin{abstract}
We prove that a system of discrete 2D in-plane
dipoles with four possible orientations,
interacting via a 3D dipole-dipole interaction plus a nearest neighbor
ferromagnetic term, has periodic striped ground states. 
As the strength of the ferromagnetic term is increased,
the size of the stripes in the ground state increases, becoming infinite, 
\ie, giving a ferromagentic ground state, when the ferromagentic interaction 
exceeds 
a certain critical value. We also give a rigorous proof of the
{\it reorientation transition} in the ground state of a 2D system of discrete
dipoles with six possible orientations, interacting via a 3D dipole-dipole
interaction plus a nearest neighbor antiferromagnetic term.
As the strength of the antiferromagnetic term is increased the ground state
flips from being striped and in-plane to being staggered and out-of-plane.
An example of a rotator model with a sinusoidal ground state is also discussed.
\end{abstract}

\maketitle

\renewcommand{\thesection}{\arabic{section}}

%%%%%%%%%%%%%%%%%%%%%%%%%%%%%%%%%%%%%%%%%%%%%%%%%%%%%%%%%%%%%%%%%%%%%%%%%%
%%%%%%%%%%%%%%%%%%%%%%%%%%%%%%%%%%%%%%%%%%%%%%%%%%%%%%%%%%%%%%%%%%%%%%%%%%
\section{Introduction}
\setcounter{equation}{0}
%%%%%%%%%%%%%%%%%%%%%%%%%%%%%%%%%%%%%%%%%%%%%%%%%%%%%%%%%%%%%%%%%%%%%%%%%%
%%%%%%%%%%%%%%%%%%%%%%%%%%%%%%%%%%%%%%%%%%%%%%%%%%%%%%%%%%%%%%%%%%%%%%%%%%

Recent advances in film growth techniques and in the experimental
control of spin-spin interactions have revived interest in the low
temperature physics of thin films
\cite{JV94,P94,PKH90,BDLMPPZ07,SC93,MS92,
  BF90,MB96,SK04,SK06,EK93,TKNV05}.  These quasi-2D systems show a
wide range of ordering effects including formation of striped states,
reorientation transitions, bubble formation in strong magnetic fields,
etc. \cite{AB92,SW92,KP93}.  The origins of these phenomena are, in
many cases, traced to competition between short ranged exchange
(ferromagnetic) interactions, favoring a homogeneous ordered state,
and the long ranged dipole-dipole interaction, which opposes such
ordering on the scale of the whole sample.  The present theoretical
understanding of these phenomena is based on a combination of
variational methods and a variety of approximations, \eg, mean-field
and spin-wave theory \cite{DMW00,GTV00,LEFK94,KP93,MWRD95}. The
comparison between the predictions of these approximate methods and
the results of MonteCarlo simulations are often difficult, because of
the slow relaxation dynamics associated with the long-range nature of
the dipole-dipole interactions \cite{DMW00,TKNV05}.  It would clearly
be desirable to have more rigorous results about the spontaneous
formation of such patterns.  In a previous paper \cite{GLL06} we began
to investigate these questions by means of a spin-block
reflection-positivity method which, combined with apriori estimates on
the Peierls' contours, allowed us to:

{\it (i)} Describe the zero temperature phase diagram of a 1D Ising
model with nearest neighbor ferromagnetic and long range, reflection positive,
antiferromagnetic interactions. These include power-law type interactions, 
such as dipolar-like interactions. We
proved in particular the existence of a sequence of phase transitions
between periodic states with longer and longer periods as the nearest
neighbor ferromagnetic exchange strength $J$ was increased.

{\it (ii)} Derive upper and lower bounds on the ground state energy of a
class of 2D Ising models with similar competing interactions,
which agreed within exponentially small terms in $J$
with the energy of the striped state.

In this paper we prove for some models of dipole systems with discrete
orientations, on a 2D lattice, that their
ground states {\it display periodic striped order}.
As in the 1D case the stripe size increases with the strength of the nearest
neighbor exchange interaction, becoming infinite, \ie, giving a ferromagentic
ground state, when the ferromagentic interaction exceeds a certain critical
value. The proof is again based on a combination
of reflection-positivity and Peierls' estimates and on an exact reduction
of our 2D model to the 1D Ising model studied in \cite{GLL06}. The analysis
takes explicit account of the tensorial nature of the 3D dipole-dipole
interaction that, in the {\it absence} of any short range exchange, tends to
produce order in the form of
polarized columns (or rows) of aligned spins, with alternating polarization.
For dipoles oriented along four possible directions
at each site the ground state shown in Fig.1 is $4$-fold degenerate 
(spin reversal
and $90^o$ rotations) and these are the {\it only} ground states.
See \cite{DMW00} for a more detailed description of  pure dipole
states and \cite{FS81} for some  proofs.

\begin{figure}[ht]
\hspace{1 cm}
\includegraphics[height=3.5 in]{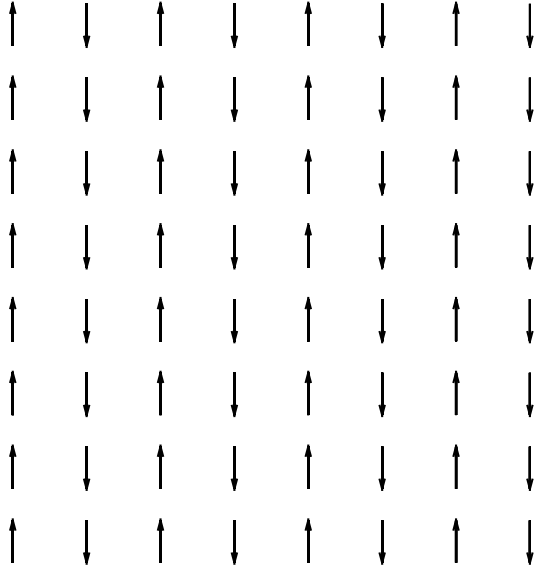}
\vspace{-0.8truecm}
\caption{A ground state of the pure dipole-dipole system in 2D.}
\end{figure}

The order described in Fig.1 is induced by the fact that
two dipoles with the same orientation attract
if their axes are parallel to their relative position
vector, and they repel if both their axes are perpendicular to their relative
position vector: in other words the dipole-dipole interaction, although overall
antiferromagnetic (in the sense that it prefers total spin equal to zero) is,
roughly speaking, ferromagnetic (FM) in one direction and antiferromagnetic in
the orthogonal direction. It is then possible to show that, in the presence
of an additional nearest neighbor ferromagnetic exchange, the FM order
will persist in one direction. In the orthogonal direction the column-column
interaction can then be effectively described in terms of a 1D model, which can
be treated by the methods of \cite{GLL06}.

A similar method allows us to investigate the zero temperature phase
diagram of the 2D discrete dipole model in the presence of a nearest
neighbor {\it antiferromagnetic} exchange, even in the case that the
dipoles are allowed to orient along six different directions (four
in-plane and two out-of-plane). In this case we prove the so-called
{\it reorientation transition} \cite{DMW00,AWDM01}, consisting in a
flip from an in-plane ground state, like the one depicted in Fig.1, to
an out-of-plane staggered state, as the strength of the
antiferromagnetic exchange interaction is increased.

The effects of the interplay between the tensorial dipole-dipole
interaction and a short-range exchange interaction is still far from
being understood in general systems of continuous (Heisenberg-like)
spins, or in general anisotropic models with different assumptions on
the allowed spin orientations and/or on the short range exchange
interactions. It is worth remarking that in these classes of models
the naive procedure of minimizing the Fourier transform of the pair
interaction not only is wrong (because it neglects the local
constraints coming from the requirement that the norm of the spin
vector at each site is equal to 1) but generally leads to very bad
estimates on the ground state energy and on the size of the
characteristic zero temperature patterns. However it is interesting to
note that, for an $O(n)$ spin model, $n\ge 2$, with a {\it scalar}
interaction whose Fourier transform admits a non trivial minimum at
$\kk_0\neq\V0$, it is relatively simple to show existence of ``soft''
striped, \ie, sinusoidal 1D spin wave, ground states. This was shown
via a very general argument by Nussinov \cite{Nussinov} for classical
$O(n)$ spins. This general result is independent of the dimension and
of any reflection-positivity of the interactions. It is applied here
to the concrete example of rotators in two dimensions interacting with
a nearest neighbor ferromagnetic interaction and a weak long range
{\it scalar} $1/r^3$ interaction, \ie, a scalar $O(n=2)$ model with
competing long range interactions, imitating the decay properties of
the real dipole-dipole potential. The fact that the solution to this
scalar isotropic problem is easy adds further motivation to the study
of the harder and more exciting case of real dipole systems.  (Note
that even the existence of the thermodynamic limit in an external
field, which is expected to be shape dependent, is unproven for a 3D
dipole system).

The rest of this paper is organized as follows. In Section \ref{sec2}
we introduce the 2D spin model with discrete orientations and state
our results about its zero temperature phase diagram. This includes
existence of striped order (in the case of a short range ferromagnetic
exchange) and existence of a reorientation transition from an in-plane
to an out-of-plane ordered state (in the case of an antiferromagnetic
exchange). In Section \ref{sec3} we discuss an example of a  2D
rotator model with long-range scalar interactions whose ground states are
given by
sinusoidal 1D spin waves (``soft stripes'').  In Sections \ref{sec4}
to \ref{secanti} we present the details of the proof for the 2D spin
model with discrete orientations, both for the case of ferromagnetic
and of antiferromagnetic exchange. In Section \ref{sec7} we summarize
some aspects of the conjectured positive-temperature behavior of
dipole systems with competing interactions. The rigorous analysis of
the phase diagram at positive temperatures is presently beyond our
reach.

Some results and proofs in the present paper rely on those of the previous
paper \cite{GLL06}. We have discovered a minor technical error in the proofs of
Theorems 1 and 2 of \cite{GLL06} which was caused by overlooking several
exponentially small terms of the form $e^{-cL}$ which came from the use
of periodic boundary conditions on a ring of length $L$.
These errors can be repaired, but we found
that everything can be done more easily and clearly using open boundary
conditions on the line rather than periodic boundary conditions on the circle.
This improved methodology, which might be independently interesting for future
work, is given here in Appendix \ref{A}.

%%%%%%%%%%%%%%%%%%%%%%%%%%%%%%%%%%%%%%%%%%%%%%%%%%%%%%%%%%%%%%%%%%%%%%%%%%
%%%%%%%%%%%%%%%%%%%%%%%%%%%%%%%%%%%%%%%%%%%%%%%%%%%%%%%%%%%%%%%%%%%%%%%%%%
\section{2D discrete dipoles: the model and the Main results}
\setcounter{equation}{0}\label{sec2}
%%%%%%%%%%%%%%%%%%%%%%%%%%%%%%%%%%%%%%%%%%%%%%%%%%%%%%%%%%%%%%%%%%%%%%%%%%
%%%%%%%%%%%%%%%%%%%%%%%%%%%%%%%%%%%%%%%%%%%%%%%%%%%%%%%%%%%%%%%%%%%%%%%%%%

In this section we introduce the 2D model with discrete orientations and state
the main results on the structure of its zero temperature phase diagram.
We shall first discuss the stripe formation phenomenon in the presence of a
nearest neighbor ferromagnetic interaction and then the reorientation
transition phenomenon in the presence of an antiferromagnetic exchange
interaction.

\subsection{The ferromagnetic case}\label{sec2A}
Let $\L\subset \ZZZ^2$ be a simple cubic 2D torus of side $2L$ and let
$\vec S_\xx$, $\xx\in\L$, be an in--plane unit vector with components
$\{S^i_\xx\}_{i=1,2}$. We shall assume that $\vec S_\xx$ can only be oriented
along the two coordinate directions of $\L$ (\ie, that
$S^1_\xx$ and $S^2_\xx$ can take values $\{-1,0,1\}$,
with $(S^1_\xx)^2+(S^2_\xx)^2=1$). We shall denote by $\O_\L$ the
corresponding spin configuration space and, for later convenience, we shall
define $\O^V_\L=\{\vec{\ul S}_\L\in\O_\L : S^1_\xx=0,\forall \xx\in\L\}$
and $\O^H_\L=\{\vec{\ul S}_\L\in\O_\L : S^2_\xx=0,\forall \xx\in\L\}$ to
be the subspaces of vertical and horizontal spin configurations.
The Hamiltonian is of the form:
\be H=\sum_{i,j=1}^2\sum_{\xx,\yy\in\L}
S^i_\xx W_{ij}(\xx-\yy) S^j_\yy-\sum_{<\xx,\yy>\in\L}
\left[J\vec S_\xx \cdot\vec S_\yy +\l\big((\vec S_\xx \cdot\vec S_\yy)^2-1
\big)\right]
\label{2.1}\ee
where, denoting the Yukawa
potential by $Y_\e(\xx)=e^{-\e|\xx|}|\xx|^{-1}$, the interaction matrix
$W(\xx)$ is of the dipole form given by:
\be W_{ij}(\xx)=\sum_{\Vn\in\zzz^3}(-\dpr_i\dpr_j)Y_\e(\xx+2\Vn L)\;, \qquad
\xx\neq\V0\label{1.2}\ee
and $W_{ij}(\V0)=\sum_{\Vn\neq\V0}(-\dpr_i\dpr_j)
Y_\e(2\Vn L)$. The second sum in (\ref{2.1}) runs over pairs of
nearest neighbor sites in $\L$ and the constants $J$ and $\l$
will be assumed nonnegative. The $\l$ term is inserted in $H$ to
discourage neighboring spins from having orthogonal polarizations. It has
the effect of encouraging stripes, but this term alone cannot create stripes.
Without the $J$ term the ground state would be as in figure 1.

One of the main result of this section concerns the zero temperature
phase diagram of model (\ref{2.1}) with $\e=0$, that is
in the case that the long range dipole-dipole interaction is the ``real''
dipolar one. It is summarized in the following Theorem.
\\

{\bf Theorem 1.} {\it Let $\e=0$ and $0\le J< J_0(0)$,
where
\be J_0(0)=\sum_{m\in\zzz}\int_{-\io}^\io\,dk\,
\frac{\p^2m^2}{\sqrt{4\p^2m^2+k^2}}\left(\sinh\frac{\sqrt{4\p^2m^2+k^2}}2
\right)^{-2}\;.\label{2.4aa}\ee
There exists $\l_0(J)$ such that, if $\l\ge \l_0(J)$, then the
specific ground state energy of (\ref{2.1})
in the thermodynamic limit is given by:
\be \lim_{|\L|\to\io}\fra1{|\L|} E_0(\L)=\min_{h\in\zzz^+}e(h)\label{2.5}\ee
where $e(h)\defin\lim_{|\L|\to\io} |\L|^{-1}E_{per}^{(h)}(\L)$ and
$E_{per}^{(h)}(\L)$ is the energy of a periodic configuration with either
vertical or horizontal stripes all of size $h$ and alternate magnetization.
If the side of $\L$ is divisible by the optimal period (\ie, by $2h^*(J)$,
with $h^*(J)$ the minimizer of the r.h.s. of (\ref{2.5})) then the
{\underline{only}}
ground states are the periodic configuration with either
vertical or horizontal stripes all of size $h^*(J)$
and alternate magnetization.

If $\l=+\io$, \ie, if we consider model
(\ref{1.2}) restricted to $\O^V_\L\cup\O^H_\L$, then the same conclusions are
valid for all $J\ge 0$.}\\
\\
{\bf Remarks.} 1)
For any $J< J_0(0)$, the ground state has non trivial stripes of finite size,
that is the minimizer $h^*(J)$ defined in the Theorem is finite, and
$h^*(J)$ diverges logarithmically as $J$ tends to $J_0(0)$.
In the hard core ($\l=+\io$) case, Theorem 1 implies that for any $J\ge J_0(0)$
we have $h^*(J)=+\io$ and the ground state is ferromagnetic.
\\
2) The constant $\l_0(J)$ introduced in the Theorem is proportional to
$h^*(J)$, \ie, it is of the form $\l_0(J)=C\, h^*(J)$, with $C$ independent
of $J$. In particular $\l_0(J)$ diverges at $J=J_0(0)$.
It is not clear whether this divergence is an artifact of our proof
and whether one should expect the same result to be valid for smaller
values of $\l$ (in particular for $\l=0$). For $\l=0$,
we tried to look for states with all four possible orientations and
energy smaller than the one given by (\ref{2.5}), but we did not succeed.
It could very well be that the same result is actually valid for any
$\l\ge 0$, but unfortunately we don't know how to prove or disprove it.\\

In the case that $\e>0$, we can extend the previous result to all values
of $J\ge 0$ and finite $\l$.
The result is summarized in Theorem 2.\\

{\bf Theorem 2.} {\it The conclusions of Theorem 1 are also valid if:
$\e>0$, $J\ge 0$ and $\l\ge\max\{C_\e-J,0\}$, where
\be C_\e=\const.\,
\cases{|\log\e| & if $\ \e\le 1/2\;,$ \cr
\e e^{-\e} & if $\ \e>1/2\;.$ \cr}
\label{ceps}\ee}
\\
\\
{\bf Remarks.}
1) Similarly to the $\e=0$ case, the optimal stripe size $h^*(J)$ is finite if
and only if $J<J_0(\e)$, where
\be J_0(\e)=\sum_{m\in\zzz}\int_{-\io}^\io\,dk\,
\frac{\p^2m^2}{\a(m,k,\e)}\left(\sinh\frac{\a(m,k,\e)}2\right)^{-2}
\;,\label{2.4}\ee
with
\be \a(m,k,\e)=\sqrt{4\p^2m^2+k^2+\e^2}\;,\label{alpha}\ee
The constant in the r.h.s. of (\ref{ceps}) is chosen such that $C_\e$ is
larger than $J_0(\e)$,
for all $\e>0$. Note that the conclusions of Theorems 1 and 2 are also valid
under the assumptions: $\e>0$, $J\ge 0$ and $\l\ge \const\, h^*(J)$.
\\
2) For small $\e$,
the bound on the value of $\l$ above which we get
the striped state is proportional to $|\log\e|$ and in
particular diverges at $\e=0$. Also in this case, it is not clear whether this
divergence is just an artifact of our proof
and whether one should expect the same result to be valid for smaller
values of $\l$.\\

In summary, we prove that the ground state of (\ref{2.1}) is a
periodic array of stripes if the $\l$ term is large enough,
and more precisely: in the dipole case, $\e=0$, if $\l$ is larger than a
constant
proportional to the period of the $\l=+\io$ ground state; in the Yukawa
case, $\e>0$, if $\l$ is larger than a constant depending on the
Yukawa mass (and diverging logarithmically as it goes to zero).
The proof of Theorems 1 and 2 in the $\l=+\io$ case,
described in Section \ref{sec4}
below, is based on an exact reduction to an effective 1D model, to be treated
by the methods of \cite{GLL06}: as a byproduct of the proof, we also get
explicit bounds for the energies of excited states.
The extension to finite $\l$, both in the $\e=0$ and $\e>0$ case,
is based on a Peierls' estimate on the energy
of a droplet of horizontal spins surrounded by vertical spins.
The proof of the Peierls' estimate is very simple under the assumptions of
Theorem 2, that is in the presence of
exponential decay with mass $\e$ and of a
large penalty $\l$ (large non uniformly in the mass $\e$); this proof, together
with the necessary definitions of Peierls' contours and droplets, is
described in Section \ref{sec5}. Extending
the result to the case $\e= 0$ and $\l$ finite (\ie, the case considered in
Theorem 1) is highly non-trivial: in fact, since the long range potential
decays as the third power of the distance, the naive dimensional estimate on
the dipole energy of a droplet of size $\ell$ decreases as $-\ell\log\ell$.
So in order to exclude the presence of droplets of spins with
the ``wrong'' orientation we have to use cancellations, that is we have
to prove the presence of screening in the ground state. The proof
is given in Section \ref{sec6}.

\subsection{The antiferromagnetic case}\label{sec2B}
Let us now consider the case that the $\vec S_\xx$ have six possible
orientations, four in-plane and two out-of-plane, \ie, $S^1_\xx$, $S^2_\xx$ and
$S^3_\xx$ can take values $\{-1,0,1\}$, with $(S^1_\xx)^2+(S^2_\xx)^2+
(S^3_\xx)^2=1$. Their interaction is given by the Hamiltonian (\ref{2.1}), with
$\xx$ still located at the sites of the 2D torus $\L$ considered above,
with $J\le 0$ and $\e=0$. Then,\\

{\bf Theorem 3.} {\it There exists an absolute constant $\l_0\ge 0$ such that,
if $\l\ge \max\{\l_0-|J|,0\}$, then the specific ground state energy
is given by:
\be\lim_{|\L|\to\io}\frac1{|\L|}E_0(\L)=\min\{e_1,e_0-2|J|\}
\label{AFspecific}\ee
with $e_1$ the specific energy of the planar antiferromagnetic state
described in Fig.1 and $e_0-2|J|$ the specific energy of the out-of-plane
staggered state. One has $e_0>e_1$.
For $|J|<(e_0-e_1)/2$ and $\L$ large enough the planar antiferromagnetic
states (\ie, the one described in Fig.1 plus those obtained from it by
translation and/or by a $90^o$ rotation) are the \underline{only}
ground states. For
$|J|>(e_0-e_1)/2$ and $\L$ large enough the out-of-plane N\'eel states
are the \underline{only} ground states.
For $|J|=(e_0-e_1)/2$ and $\L$ large enough
the out-of-plane N\'eel state and the in-plane state
described in Fig.1 are the \underline{only} ground states.}\\
\\
Theorem 3 is proven in Sec.\ref{secanti}. Its proof goes along the lines
of the proof of Theorem 2.
In other words, we first consider the restricted problem with the dipoles
all constrained to be parallel
(either in- or out-of-plane), and determine the ground states using a
reflection positivity argument. Then we prove that for $\l$ large enough
the picture doesn't change, using a Peierls' argument.
Note that now the critical value $\l_0$ is an absolute constant,
independent of $J$: the Theorem can be extended to cover the $\e>0$
case, in which case $\l_0$ is also independent of $\e$. This is due to the
fact that the reference states we now need to consider
(\ie, the ground states of the restricted problem) are always
antiferromagnetic with period two, and this induces a strong screening
which effectively corresponds to a faster decay of interactions, at least
as the inverse distance to the power four.

%%%%%%%%%%%%%%%%%%%%%%%%%%%%%%%%%%%%%%%%%%%%%%%%%%%%%%%%%%%%%%%%%%%%%%%%%%
%%%%%%%%%%%%%%%%%%%%%%%%%%%%%%%%%%%%%%%%%%%%%%%%%%%%%%%%%%%%%%%%%%%%%%%%%%
\section{$O(n)$ spins with scalar interactions}
\setcounter{equation}{0}\label{sec3}
%%%%%%%%%%%%%%%%%%%%%%%%%%%%%%%%%%%%%%%%%%%%%%%%%%%%%%%%%%%%%%%%%%%%%%%%%%
%%%%%%%%%%%%%%%%%%%%%%%%%%%%%%%%%%%%%%%%%%%%%%%%%%%%%%%%%%%%%%%%%%%%%%%%%%

In this section we describe an example of a spin system
with dipole-type interactions that displays periodic striped ground states.
The construction of the ground state, based on an observation about
$O(n)$ spin models with {\it scalar} interactions, was originally
given by Nussinov \cite{Nussinov}. This shows that striped states are, 
very naturally,
the ground states for a large class of $O(n)$ models with $n\ge 2$ and
competing interactions. For this class of models the local constraint
on the norm of the spin at each site is not strong enough to
invalidate the naive procedure for determining the ground states based
on minimization of the Fourier transform of the interaction.

Let us consider an $O(n)$ model, $n\ge 2$, on a simple cubic $d$--dimensional
torus $\L\subset\ZZZ^d$ of side $L$ with $d\ge 1$ and Hamiltonian:
\be H=\sum_{\xx,\yy}J(\xx-\yy)\,\vec S_\xx\cdot\vec S_\yy\label{01.1}\ee
where $J(\xx)$ is a function of the distance $|\xx|$ only and the $\vec S_\xx$
are $n$-dimensional unit vectors, $n\ge 2$.
The Hamiltonian (\ref{01.1}) can be rewritten in Fourier space as follows:
\be H=\sum_{i=1}^n\sum_{\kk\in \DD_L}
\hat J(\kk)\,\hat S_\kk^i\cdot\hat S_{-\kk}^i\label{01.2}\ee
where $\DD_L=\{\kk=2\p L^{-1}\Vn\;,\ \Vn\in[0,L)^d\cap\ZZZ^d\}$ and
\be \hat S_\kk^i=\frac1{\sqrt{|\L|}}\sum_\xx S_\xx^i e^{i\xx\cdot\kk}\,,
\qquad \hat J(\kk)=\sum_\xx J(\xx)e^{i\xx\cdot\kk}\label{01.3}\ee

{\bf Proposition 1 (Nussinov).} {\it
If $\kk_0$ is a minimizer for $\hat J(\kk)$ and $g\in O(n)$ is an $n\times n$
orthogonal matrix, then all states of the form
\be \vec S^{\;g,\kk_\V0}_\xx=g\; \vec S^{\kk_\V0}_\xx\;, \qquad{\rm with}
\qquad \vec S^{\kk_\V0}_\xx=
\big(\cos(\kk_\V0\cdot\xx)\,,\,\sin(\kk_\V0\cdot\xx)
\,,\, 0\,,\, \cdots\,,\,0\big)\;,\label{01.4}\ee
are ground states of model (\ref{01.2})-(\ref{01.3}).}\\

{\cs Proof.}
By (\ref{01.2}) and the normalization condition
$\sum_{i,\kk}|\hat S_\kk^i|^2=|\L|$ we see
that the ground state energy cannot be smaller than $|\L|\min_\kk\hat J(\kk)$;
on the other hand the states (\ref{01.4}) all have energy equal to
$|\L|\min_\kk\hat J(\kk)$ and they
satisfy the normalization $|\vec S_\xx|^2=1$, so this proves that
they are ground states.\qed

\vspace{.3truecm}

Note that if $\kk_0\neq\V0$,
the ground states $\{\vec S^{\;g, \kk_\V0}_\xx\}_{\xx\in\L}$ are ``striped'',
in the sense that they are 1D sinusoidal spin waves in the
direction $\kk_\V0/|\kk_\V0|$ with wavelength $2\p/|\kk_\V0|$;
we call these {\it soft stripes}.
Under additional conditions on the geometry of the set of minima of
$\hat J(\kk)$ one could actually prove that such states are the only possible
ground states, but we will not investigate this question in the
greatest possible generality. Instead, as an illustration, we will discuss an
explicit example where these ideas can be used to infer that all ground states
are soft stripes. A similar analysis can be performed to show that the
previous remark also applies to the case of a 3D $O(n)$ model, $n\ge 2$, with
spins interacting via a short range ferromagnetic interaction plus a positive
long range Coulombic $1/r$ interaction, \ie, to the class of models considered
in \cite{CEKNT96}.

\\
{\bf Example.} {\it Let $\L\subset \ZZZ^2$ be a simple cubic 2D torus
of side $L$ and let $H$ be defined as
\be H=-J\sum_{<\xx,\yy>}\vec S_\xx\cdot\vec S_\yy+
\e\sum_{\mm\in\zzz^2}\sum_{\xx,\yy\in\L\atop \xx\neq
\yy}\frac{\vec S_\xx\cdot\vec S_\yy}{|\xx-\yy+\mm L|^3}\label{01.5}\ee
where the first sum ranges over nearest neighbor sites of $\L$ and the
$\vec S_\xx$'s are 2D unit vectors. Then there exists
a constant $c>0$ such that, if $\e\le c J$ and $L$ is large enough,
\underline{all}
the ground states of $H$ are soft striped states.}\\
\\
{\bf Remark.} This result, as well as the general remark above,
depends crucially on the fact that the interaction is scalar and
{\it isotropic}, \ie, for each pair $(\xx,\yy)$, the coupling depends only on
$\vec S_\xx\cdot\vec S_\yy$. In the case of anisotropic interactions,
the ground state may look very different, and will be in general very
difficult to identify. As an illustrative example, take the case of soft scalar
spins, \ie, scalar spins $S^1_\xx$ constrained to satisfy $|S^1_\xx|\le 1$,
interacting via a pair potential $J(\xx)$ as in (\ref{01.1}), with the further
property that $J(\V0)=0$.
This example can be viewed as an extreme anisotropic case,
where all couplings between the $i$-th components of the spins, $i\neq 1$, have
been switched off. In this case the ground state will be of the Ising type;
this can be shown in the following way. The Hamiltonian is linear in
$S^1_\xx$, for each $\xx\in\L$, and it has to be minimized under the
constraint that $|S^1_\xx|\le 1$, for all $\xx\in\L$. By the aforementioned
linearity, the minimum will be attained at the boundary, that is
$S^1_\xx=\pm1$, for all $\xx\in\L$. The problem then reduces to determining
the ground state of an Ising model with both long range antiferromagnetic and
nearest neighbor ferromagnetic interaction, that is a very difficult and,
in many respects, open problem, see \cite{GLL06,MWRD95}.
For intermediate values of the anisotropy the problem is even more
difficult and it is unclear whether the transition
from a sinusoidal spin wave state to an Ising-like state as the
couplings between the $i$-th components of the spins, $i\neq 1$, are
decreased will be a sharp transition or rather a continuous one.
\\

{\cs Proof of the Example.} The Hamiltonian (\ref{01.5}), up to an overall
constant, can be rewritten as in (\ref{01.2}), with $\hat J(\kk)=
2J\sum_i(1-\cos k_i)+\e\sum_{\xx\neq\V0}e^{i\kk\xx}|\xx|^{-3}$.
Note that $\sum_{\xx\neq\V0}e^{i\kk\xx}|\xx|^{-3}$ can be conveniently
rewritten as $\hat g(\kk)-1+\sum_{\pp\in 2\p\zzz^2}\int d\xx e^{i(\pp+\kk)\xx}
(\xx^2+1)^{-3/2}$, where $\hat g(\kk)$ is the Fourier transform of
$g(\xx)=(1-\d_\xx)\big[|\xx|^{-3}-(\xx^2+1)^{-3/2}\big]$
(note that $g(\xx)$ goes to zero as $|\xx|^{-5}$ as $|\xx|\to\infty$).
Using the fact that $\int d\xx e^{i(\pp+\kk)\xx}(\xx^2+1)^{-3/2}=2\p
e^{-|\pp+\kk|}$ (see \cite{GR00}), we find that
\be \hat J(\kk)=2J\sum_{i=1}^2(1-\cos k_i)+2\p\e
\sum_{\pp\neq\V0}e^{-|\pp+\kk|}+\e\hat g(\kk)-\e\label{01.6}\ee
where $\hat g(\kk)$ is twice differentiable and even in $\kk$. An elementary
study of the minima of (\ref{01.6}) shows that for $\e/J$ small
they are located within $O(\e/J)^2$ from the points $(\pm\p\e/J,0)$,
$(0,\pm\p\e/J)$. Moreover the minima have the following property
(that we shall call {\it non-degeneracy}):
there are no two distinct (unordered) pairs of minimizing vectors $(\pp,\qq)$,
$(\pp',\qq')$, such that $\pp+\qq=\pp'+\qq'$; this is because the minima
all lie on a surface of strictly positive curvature. This implies that
all ground states are ``striped'', that is they are all of the form
(\ref{01.4}), with stripes that are all (almost) horizontal or vertical.
In fact one can show that no state obtained as a
superposition of different minimizing modes can be a ground state: this is
proven by using the normalization $|\vec S_\xx|^2=1$ and
the non-degeneracy condition. For a similar discussion, 
see \cite{Nussinov}. \qed

%%%%%%%%%%%%%%%%%%%%%%%%%%%%%%%%%%%%%%%%%%%%%%%%%%%%%%%%%%%%%%%%%%%%%%%%%%
%%%%%%%%%%%%%%%%%%%%%%%%%%%%%%%%%%%%%%%%%%%%%%%%%%%%%%%%%%%%%%%%%%%%%%%%%%
\section{2D discrete dipoles: the $\l=+\io$ case}
\setcounter{equation}{0}\label{sec4}
%%%%%%%%%%%%%%%%%%%%%%%%%%%%%%%%%%%%%%%%%%%%%%%%%%%%%%%%%%%%%%%%%%%%%%%%%%
%%%%%%%%%%%%%%%%%%%%%%%%%%%%%%%%%%%%%%%%%%%%%%%%%%%%%%%%%%%%%%%%%%%%%%%%%%

As mentioned in Section \ref{sec2} we shall first  prove
Theorems 1, 2 and 3 in the case $\l=+\io$, that is in the case that
the spins are either all vertical or all horizontal or, possibly,
all out-of-plane. Then we will show that, if $\l$ is large enough, the spins in
the ground state configurations will in fact be either all vertical or all
horizontal. We will do this first for the
ferromagnetic plus Yukawa case (Section \ref{sec5}),
then for the ferromagnetic plus dipole case (Section \ref{sec6}) and finally
for the antiferromagnetic plus dipole case (Section \ref{secanti}).

\subsection{The ferromagnetic case.}\label{sec4A}
Let us assume here that $\e,J\ge 0$ in (\ref{2.1}) and (\ref{1.2})
and the spins are oriented along four possible directions, as described in
Section \ref{sec2A}. Let us denote by $H_\io$ the Hamiltonian (\ref{2.1}) with
the hard core interaction corresponding to $\l=+\io$. By definition, $H_\io$
is the restriction of (\ref{2.1}) to $\O^V_\L\cup\O^H_\L$, \ie to the
space of configurations with spins either all vertical
or all horizontal. Without loss of generality,
let us consider the case that all spins are vertical, that is $\vec S_\xx=(0,
\s_\xx)$, $\s_\xx=\pm1$, $\forall \xx\in\L$. In the following we shall
alternatively refer to the spins as ``up or down'' spins or ``plus and minus''
spins, with interchangeable meaning. The key remark is that in this
case the system is reflection positive with respect to ferromagnetic
reflections in horizontal lines. The relevant reflection symmetry is defined as
follows. Let $\p$ be a {\it pair} of horizontal lines midway
between two lattice rows which bisect the torus $\L$ of side $2L$ into two
pieces $\L_+$ and $\L_-$ of equal size. Let $r$ denote reflection of sites with
respect to $\p$. Clearly $r\L_-=\L_+$. We define
\be \th\s_\xx=\s_{r\xx}\label{3.1}\ee
For any function
$F(\{\s_\xx\}_{\xx\in\L})$, we shall define the reflected function $\th F$
as $\th F(\{\s_\xx\}_{\xx\in\L})=[F(\{\th\s_\xx\}_{\xx\in\L})]^*$.
Note
that if $F_+$ depends only on the spins in $\L_+$, then $F_-=\th F_+$ will
depend only on the spins in $\L_-$. We shall say that $H_\io$ is reflection
positive (RP) with respect to reflections in horizontal planes if it can be
written in the form $H_\io=
H_++\th H_+ -\int C_+(x)\th C_+(x)d\r(x)$ for a positive measure $d\r(x)$,
with $H_+,C_+(x)$ depending only on the spins in $\L_+$. In our case this
representation can be achieved by defining $H_+$ as the interaction of the
spins in $\L_+$ among themselves, $H_-=\th H_+$ as the interaction of the
spins in $\L_+$ among themselves and by suitably rewriting the interaction term
$\sum_{\xx\in\L_+}\sum_{\yy\in\L_-}\s_\xx\s_\yy (W_{22}(\xx-\yy)-J
\d_{|\xx-\yy|,1})$, in the following way.

Let $x_2>0$ and let us rewrite
\be Y_\e(\xx)=\int \frac{d\kk}{(2\p)^3} \frac{4\p}{\kk^2+\e^2}e^{i\kk\xx}=
\frac1{2\p}\int\frac{d\kk_\perp}{\sqrt{\kk_\perp^2+\e^2}}
\,e^{ik_1x_1}e^{-|x_2|\sqrt{\kk_\perp^2+\e^2}}\label{3.2}\ee
where in the last expression $\kk_\perp=(k_1,k_3)$.
If $x_2>y_2$ and $\dpr_2=\frac{\dpr}{\dpr x_2}$:
\bea&& \sum_{\xx\in\L_+\atop\yy\in\L_-}\s_\xx\s_\yy W_{22}(\xx-\yy)
=-\sum_{\Vn\in\zzz^2}\sum_{\xx\in\L_+\atop\yy\in\L_-}
\s_\xx\s_\yy \dpr_2^2 Y_\e(\xx-\yy+2\Vn L)
=\nn\\
&&=-\sum_{\Vn\in\zzz^2}\frac1{2\p}\int d\kk_\perp\,\sqrt{\kk_\perp^2+\e^2}
\,\sum_{\xx\in\L_+\atop\yy\in\L_-}\s_\xx\s_\yy e^{ik_1(x_1-y_1)}
e^{-(x_2-y_2)\sqrt{\kk_\perp^2+\e^2}}\label{3.3}\eea
that has clearly the correct structure $-\int C(x)\th C(x)d\r(x)$.
Note that the exchange interaction between $\L_+$ and $\L_-$ admits a
representation of the form $-J\sum_i\s_i\th \s_i$, with $\s_i$ the spins
in $\L_+$ at distance 1 from $\L_-$. This concludes the proof that $H_\io$
is RP w.r.t. reflections in horizontal planes.

Let $\RR_i$ be the $i$--th row of $\L$ and, given a spin configuration
$\ul\s_\L=\{\s_\xx\}_{\xx\in\L}$, let $\SS_i(\ul\s_\L)$ be the spin
configuration $\ul\s_\L=$ restricted to $\RR_i$. By reflection positivity and,
more specifically, by the chessboard estimate
(see Theorem 4.1 in \cite{FILS1}),
the energy of $\ul\s_\L$ can be bounded below as
\be H_\io(\ul\s_\L)\ge \frac1{2L}\sum_{i=1}^{2L}H_\io(\{\SS_i(\ul\s_\L),
\ldots,\SS_i(\ul\s_\L)\})\label{3.4}\ee
where $\{\SS_i(\ul\s_\L),
\ldots,\SS_i(\ul\s_\L)\}$ is the spin configuration in $\L$ obtained by
repeating in
every row the same spin configuration $\SS_i(\ul\s_\L)$.
Then the problem of minimizing
$H_\io$ among all possible spin configurations is reduced to the problem of
minimizing $H_\io$ among the spin configuration in which
all spins in a given column have the same direction. Let us consider one such
configuration. If column $x_1$ has all spins pointing up, then we will label
it by $\s_{x_1}=+$, if it has all spins pointing down we will label it by
$\s_{x_1}=-$.
The energy of the configuration
labelled by $\{\s_x\}_{x=1,\ldots,2L}$ is
\be E=2L\left[w_0-J+\sum_{x\neq y=1}^{2L}\s_xw_{\uparrow\uparrow}
(x-y)\s_y-J\sum_{x=1}^{2L}\s_x\s_{x+1}
\right]\label{3.5}\ee
with $w_0=\sum_{n\ge 1}W_{22}(n\hat e_2)$ and
\bea w_{\uparrow\uparrow}
(x-y)&&=\sum_{\Vn\in\zzz^2}W_{22}\left((x-y+2Ln_1)\hat e_1+n_2\hat e_2
\right)
=\nn\\
&&=\frac1{2\p}\sum_{\Vn\in\zzz^2}\int\frac{dk_2dk_3}{\sqrt{k_2^2+k_3^2+\e^2}}
\,k_2^2\,e^{ik_2n_2}e^{-|x-y+2Ln_1|\sqrt{k_2^2+k_3^2+\e^2}}
\label{3.6}\eea
where $\hat e_1,\hat e_2$ are the two coordinate unit vectors and
in the second line we used the representation (\ref{3.2})--(\ref{3.3}).
Performing the summations over $n_1,n_2$ we get
\be w_{\uparrow\uparrow}
(x-y)=\sum_{m\in\zzz}\int dk\frac{4\p^2m^2}{\a(m,k,\e)}
\frac{e^{-|x-y|\a(m,k,\e)}+e^{-(2L-|x-y|)\a(m,k,\e)}}{1-e^{-2L\a(m,k,\e)}}\;,
\label{3.7}\ee
where $\a(m,k,\e)$ was defined in (\ref{alpha}).
This potential decays exponentially at large distances and is reflection
positive with respect to the antiferromagnetic reflections. The ground states
of this one-dimensional model can be studied by the same methods of
\cite{GLL06}. As mentioned in the Introduction, the proofs of Theorems 1 and 2
in \cite{GLL06} actually contains a minor technical error, which was caused by
overlooking exponentially small terms which came from periodic boundary
conditions. While fixing the error, we discovered that the same results can be
proven by a simpler and more elegant method, which is presented here in the
Appendix. The result is that the ground states for the one dimensional system
described by the effective Hamiltonian (\ref{3.5}) consist of periodic
arrays of blocks of alternating sign, all of size $h$, with $h$ the
positive integer minimizing the function:
\be e(h)=w_1-2J+\frac{2J}h-\frac2h\sum_{m\in\zzz}\int_{-\io}^\io\,dk\,
\frac{4\p^2m^2}{\a(m,k,\e)}\frac{e^{-\a(m,k,\e)}}{
\big(1-e^{-\a(m,k,\e)}\big)^2}\tanh
\frac{h\,\a(m,k,\e)}{2}\label{3.8}\ee
where $w_1=w_0+\sum_{n\ge 1}w_{\uparrow\uparrow}
(n)$. Note that $e(h)$ is the specific energy
of a periodic configurations with blocks all of the same size and alternating
sign. An elementary study of the behavior of $e(h)$ shows that if $J\ge J_0$,
with $J_0$ defined as in (\ref{2.4}), then the ground state is
ferromagnetic. If $0\le J<J_0$ then the ground state is non trivial and $h$
is the integer part of the solution to
\bea&& J_0-J=2\sum_{m\in\zzz}\int_{-\io}^\io\,dk\,
\frac{4\p^2m^2e^{-(h+1)\a(m,k,\e)}\left(1+e^{-h\a(m,k,\e)}
+h\,\a(m,k,\e)\right)}
{\a(m,k,\e)\cdot\big(1-e^{-\a(m,k,\e)}\big)^2\cdot
\big(1+e^{-h\a(m,k,\e)}\big)^2}\label{3.10}\eea
In terms of the original dipole system this shows that
for $J\ge J_0$ the ground state is ferromagnetic, while for $0\le J<J_0$ the
ground state is striped, with stripes either all horizontal or all vertical.
The stripes have alternating orientation and their thickness varies from 1 to
$\io$ as $J$ is increased from $0$ to $J_0$. \\

The results in \cite{GLL06} and in the Appendix also imply a bound on the
energy of any given 2D state $\ul\s_\L$ {\it different} from the striped ground
state $\ul\s^*_\L$. From now on, let us assume for
simplicity that the minimization problem $\min_{h\in\zzz^+}e(h)$
is solved by a unique $h^*$ with the property that $e(h^*)
=\min_{h\in\zzz^+}e(h)=\min_{h\in\rrr^+}e(h)$.
Note that this is {\it not} a generic property: in general
$\min_{h\in\zzz^+}e(h)\neq\min_{h\in\rrr^+}e(h)$
and for some special values of
$J$ it could even happen that the minimization problem on $\ZZZ^+$ is
solved by two consecutive values $h^*,h^*+1$. However the general case can be
treated in a way completely analogous to the one described below, at the price
of slightly more cumbersome notation \footnote{
The idea for treating the problem in full
generality is to introduce the set ${\cal H}^*=\{h^*-1,h^*,h^*+1\}$
of ``optimal'' lengths -- here $h^*$ is the smallest minimizer of
the minimization problem $\min_{h\in\zzz^+}e(h)$ -- and then to distinguish
lengths of ``optimal size'' $h\in{\cal H}^*$ from those of ``wrong size''
$h\not\in{\cal H}^*$ (instead of simply distinguishing $h=h^*$ from
$h\neq h^*$).}. So let us assume that $e(h^*)=\min_{h\in\zzz^+}e(h)=
\min_{h\in\rrr^+}e(h)$ and let $c_J=\frac12
\min_{h\neq h^*}(e(h)-e(h^*))$. Let us think $\ul\s_\L$
as a collection of 1D configurations corresponding to the configurations
$\SS_i(\ul\s_\L)$ on different rows (here $i=1,\ldots,2L$ is the index
labelling different rows). Moreover, let us think each $\SS_i(\ul\s_\L)$ as a
collection of blocks $B$ of size $h_B$ and alternate magnetization.
Eq.(23) in \cite{GLL06}
implies that the energy of $\ul\s_\L$ can be bounded from below as:
\be H_\io(\ul\s_\L)\ge |\L|e(h^*)+\sum_{B: h_B\neq h^*}
2c_J h_B\label{3.11}\ee
So for any block of wrong size $h_B$ we pay a penalty $c_J h_B$ and
for any block of optimal size $h^*$ we don't pay a priori any penalty.
We can actually improve the estimate above in the case that a block of
optimal size on a given row is next to a non-optimal block of size $h_B$
on the same row:
in this case we can use the proof of Eq.(26) in \cite{GLL06} to infer that
for each such pair we pay a penalty $2d_J(h^*+h_B)$, with $d_J=\frac14
\min_{h\neq h^*}(e(\{h,h^*\})-e(h^*))$, and $e(\{h,h^*\})$ is the specific
energy of a periodic configuration with blocks of sizes $(\ldots,h,h,h^*,h^*,
h,h,h^*,h^*,\ldots)$. Combining (\ref{3.11}) with the refinement we just
discussed we find:
\be H_\io(\ul\s_\L)\ge |\L|e(h^*)+\sum_{B: h_B\neq h^*}
c_J h_B+\sum_{<B_1,B_2>:\atop h_{B_1}\neq h_{B_2}=h^*}d_J(h_{B_1}+h_{B_2})
\label{3.12}\ee
where the second sum runs over pairs of nearest neighbor horizontal blocks,
such that one of the two blocks in the pair is of optimal length.
A straightforward computation, based on the explicit expression (\ref{3.8})
of the specific energy, allows one to check that
$c_J,d_J\ge \k e^{-\a h^*}$, for suitable constants $\a,\k>0$.
This concludes the proof of Theorems 1 and 2 for the $\l=+\io$ case.

\subsection{The antiferromagnetic case.}\label{sec4B}
Let us now assume that $\e\ge 0$, $J\le 0$ and that the dipoles can have
six possible orientations, as described in Section \ref{sec2B}. In the
$\l=+\io$ case, the spins can only be oriented all parallel to each other,
either out-of-plane or vertical in-plane or horizontal in-plane.

If the spins are all oriented out-of-plane, then the Hamiltonian reduces to a
long-range antiferromagnetic Ising Hamiltonian of the form:
\be H_\io^{3}=\sum_{\xx,\yy\in\L}
S^3_\xx W_{33}(\xx-\yy) S^3_\yy+|J|\sum_{<\xx,\yy>\in\L}
S_\xx^3 S_\yy^3 \label{3.13}\ee
with $S^3_\xx=\pm1$. As a consequence of the analysis in \cite{FILS2}
(see Proof of Theorem 5.1) the ground state of the Hamiltonian (\ref{3.13})
is the usual period-2 staggered state, for any value of $|J|\ge 0$. Its ground
state energy is $e_0-2|J|$, with
\be e_0=w_0^{AF}-\sum_{m\in\zzz+\frac12}\int dk\; \frac{k^2}{\a(m,k,\e)}
\frac{e^{-\a(m,k,\e)}}{1+e^{-\a(m,k,\e)}}\;,\label{3.13a}\ee
where $w_0^{AF}=\sum_{n\ge 1}(-1)^n W_{11}(n\hat e_2)$ and
$\a(m,k,\e)$ was defined in (\ref{alpha}). In order to prove Theorem 3 in the
$\l=+\io$ case we need to compare this specific energy with the specific energy
of the best possible in-plane spin configurations.

So let us consider the case that the spins are all oriented in-plane;
we can assume without loss of generality that they are all horizontal.
(The choice of horizontal rather than vertical spins is made here in order
to keep the definition of the pair $\p$ of reflection planes the same
as in the previous subsection, see the lines preceding (\ref{3.1}).)
In this case we define the variables
$\s_\xx$ in such a way that $\vec S_\xx=(\s_\xx,0)$. Now the relevant
reflection in horizontal planes, replacing (\ref{3.1}), is
$\th\s_\xx=-\s_{r\xx}$ (note the minus sign!).
By the chessboard estimate we reduce to an expression analogue to (\ref{3.5}),
given by
\be E=2L\left[w_0^{AF}-|J|+\sum_{x\neq y=1}^{2L}\s_x
w(x-y)\s_y+|J|\sum_{x=1}^{2L}\s_x\s_{x+1}
\right]\label{3.14}\ee
with $\s_x$ now representing the direction (right or left) of the
spin in $(x,0)$, $w_0^{AF}$ the constant defined after (\ref{3.13a}) and
\be w(x-y)=-\sum_{m\in\zzz+\frac12}\int_{-\io}^{+\io}dk\; \a(m,k,\e)\,
\frac{e^{-|x-y|\a(m,k,\e)}+e^{-(2L-|x-y|)\a(m,k,\e)}}{1-e^{-2L\a(m,k,\e)}}\;.
\label{3.15}\ee
For $|J|=0$ the ground state
of (\ref{3.14}) is ferromagnetic (that is, in terms of the original 2D dipole
model, the ground state is given by Fig.1 -- rotated by $90^o$).
If $|J|$ is increased then, using the same methods of \cite{GLL06},
one can prove that the ground state has a sequence of transitions
from the ferromagnetic state to periodic states of {\it antiferromagnetic}
blocks of size $h$ with alternating staggered polarization
(of period $h$ or $2h$, depending whether $h$ is odd or even).
As an illustration, the states with $h=2,3,4$ (with periods $4,3,8$,
respectively) are given by: $(\cdots+--++-\cdots)$, $(\cdots+-++-++-+\cdots)$
and $(\cdots+-+--+-++-+-\cdots)$. For any value of $|J|$ the optimal period
$h^*(|J|)$ is obtained as usual by minimizing over $h$ the specific energy
$e^{AF}(h)$ of the states with AF blocks all of size $h$ and alternating
staggered magnetization. Here $e^{AF}(h)$ is given by
\be e^{AF}(h)= w_1^{AF}-2|J|+\frac{2|J|}h-\frac2h
\sum_{m\in\zzz+\frac12}\int_{-\io}^{+\io}dk\; \a(m,k,\e)\frac{e^{-\a(m,k,\e)}}
{(1+e^{-\a(m,k,\e)})^2}\frac{1-(-e^{-\a(m,k,\e)})^h}{1+(-e^{-\a(m,k,\e)})^h}
\label{3.16}\ee
with
$$w_1^{AF}=w_0^{AF}+\sum_{m\in\zzz+\frac12}\int_{-\io}^\io dk\,
\a(m,k,\e)\,\frac{e^{-\a(m,k,
\e)}}{1+e^{-\a(m,k,\e)}}\;.$$
It is straightforward to check that the $h^*(|J|)$ minimizing $e^{AF}(h)$
for a given value of $|J|$ is equal to $+\io$ as soon as $|J|$ is larger than
a critical value $|J_1|=\sum_m\int dk \a e^{-\a}(1+e^{-\a})^{-2}$. This means
that the minimal energy in-plane state is the staggered antiferromagnet for
$|J|\ge |J_1|$.
If we define $e^{dip}_h=e^{AF}(h)+2|J|(1-h^{-1})$,
we have $e^{dip}_1<e^{dip}_h$, for all $h>1$, and moreover $e^{dip}_1<e_0$,
with $e_0$ defined in (\ref{3.13a}). For $h>1$, the
critical value of $|J|$ for which $e^{AF}(1)=e^{AF}(h)$ is given by
$2|J^h_c|=(e^{dip}_h-e^{dip}_1)/(1-h^{-1})$.
Similarly, the critical value
of $|J|$ for which $e^{AF}(1)$ is equal to the specific energy of the
out-of-plane staggered state is given by $|J^0_c|=(e_0-e^{dip}_1)/2$.

It is now clear that in order to prove (\ref{AFspecific}) in the $\l=+\io$ case
it is enough to show that $|J^h_c|>|J^0_c|$, for all $h>1$. This simply follows
from a computation, whose details we omit. The uniqueness property stated
in Theorem 3 follows along the same lines as in the ferromagnetic case.
This concludes the proof of Theorem 3 in the $\l=+\io$ case.

%%%%%%%%%%%%%%%%%%%%%%%%%%%%%%%%%%%%%%%%%%%%%%%%%%%%%%%%%%%%%%%%%%%%%%%%%%
%%%%%%%%%%%%%%%%%%%%%%%%%%%%%%%%%%%%%%%%%%%%%%%%%%%%%%%%%%%%%%%%%%%%%%%%%%
\section{The Ferromagnetic Case: Finite $\l$ and Positive Mass.}
\setcounter{equation}{0}\label{sec5}
%%%%%%%%%%%%%%%%%%%%%%%%%%%%%%%%%%%%%%%%%%%%%%%%%%%%%%%%%%%%%%%%%%%%%%%%%%
%%%%%%%%%%%%%%%%%%%%%%%%%%%%%%%%%%%%%%%%%%%%%%%%%%%%%%%%%%%%%%%%%%%%%%%%%%

In this section we prove Theorem 2. We prove it by showing
that if $\e>0$ and $\l$ is larger then $\max\{C_\e-J,0\}$,
see (\ref{ceps}), then
in the ground state of (\ref{2.1}) the spins must be either all horizontal or
all vertical. Note that if this is the case then Theorem 2
simply follows from the discussion of previous section.
Let us recall that spins now can assume 4 possible directions (those parallel
to the two in-plane coordinate axis). We shall denote by $\ul{\vec S}_X$ spin
configurations in $X\subset\ZZZ^2$ [in particular we shall denote by
$\ul{\vec S}_\L$ spin configurations in $\L$] and
by $\vec{\ul S}^H$ and $\vec{\ul S}^V$ two (arbitrarily chosen)
$\l=+\io$ infinite volume ground state configurations with horizontal and
vertical stripes, respectively.

We need to introduce some definitions. As in the basic Peierls construction we
introduce the definitions of {\it contours} and {\it droplets}.
Given any configuration $\ul{\vec S}_\L$,
we define $\D_V=\D_V(\ul{\vec S}_\L)$
to be the set of sites at which the spins are vertical.
We draw around each $\xx\in\D_V$ the $4$ sides of the unit square
centered at $\xx$ and suppress the faces which occur twice: we obtain
in this way a {\it closed polygon} $\G(\D_V)$ which can be thought
as the boundary of $\D_V$. Each face of $\G(\D_V)$ separates a point
$\xx\in\D_V$ from a point $\yy\not\in\D_V$.
Along a vertex of $\G(\D_V)$ there can be either 2 or 4 lines meeting.
In the case of 4 lines, we deform slightly the polygon, ``chopping off''
the vertex from the cubes containing a horizontal spin.
When this is done
$\G(\D_V)$ splits into disconnected polygons $\g_1,\ldots,\g_r$ which we shall
call {\it contours}. Note that, because of the choice of periodic boundary
conditions, all contours are closed but can possibly wind around the
torus $\L$. The definition of contours naturally induces a notion of
connectedness for the spins in $\D_V$: given $\xx,\xx'\in\D_V$ we say
that $\xx$ and $\xx'$
are connected iff there exists a sequence $(\xx=\xx_0,\xx_1,\ldots,\xx_n=\xx')$
such that $\xx_m,\xx_{m+1}$, $m=0,\ldots,n-1$, are nearest neighbors and none
of the bonds $(\xx_m,\xx_{m+1})$ crosses $\G(\D_V)$.
The maximal connected components
$\d_i$ of $\D_V$ will be called $V$-{\it droplets} and the set of $V$-droplets
of $\D_V$ will be denoted by $\DD_V(\D_V)=\{\d_1,\ldots,\d_s\}$, or simply
$\DD_V$.
Note that the boundaries $\G(\d_i)$ of the $V$-droplets $\d_i\in\DD_V$
are all distinct subsets of $\G(\D_V)$ with the property: $\cup_{i=1}^s
\G(\d_i)=\G(\D_V)$. Similarly we can introduce the notion of $H$-droplets
and of the set $\DD_H(\D_H)$. A {\it droplet} will be either a $V$-droplet
or an $H$-droplet. The set of droplets will be denoted by $\DD=\DD_H\cup\DD_V$.

The same kind of construction allows one to define
FM-droplets as the maximal connected regions of spins
with the same orientation. We shall call FM-contours the boundaries of
FM-droplets.

Given the previous definitions, we can now state and prove the main results of
this section.\\
\\
{\bf Lemma 1} (Peierls' estimate - massive case - small $J$).
{\it If $\l\ge J+\overline C_\e$, with
$$\overline C_\e=\cases{c_1|\log\e| & if $\ \e\le 1/2$\cr c_2\e e^{-\e} & if$\
\e>1/2$\cr}$$
for two suitable constants $c_1,c_2>0$ (chosen in such a way that
in particular $\overline C_\e\ge J_0$, with $J_0$ the constant in (\ref{2.4}))
then the following is true.
Let $\ul{\vec S}_\L$ be a spin configuration in $\L$ and let
$\d\in\DD$ be one of its droplets. If $\d$ is a $V$-droplet (resp.
$H$-droplet),
the spin configuration $\vec{\ul T}_\L$
coinciding with $\ul{\vec S}_\L$ on $\d^c$ and with $\vec{\ul S}^H$ (resp.
$\vec{\ul S}^V$) on $\d$ satisfies $H(\ul{\vec S}_\L)
-H(\ul{\vec T}_\L)> 0$.}\\
\\
{\bf Lemma 2} (Peierls' estimate - massive case - large $J$).
{\it Let $\overline C_\e$ the same as in Lemma 1 and $\l\ge 0$.
Then the energy of any spin configuration associated to the set $\G$
of FM-contours can be bounded below by $E_{FM}(\L)+(J-\overline C_\e)
\sum_{\g\in\G}|\g|$, where $E_{FM}(\L)$ is the energy of the
ferromagnetic state and $|\g|$ is the length of the contour $\g$.}\\
\\
{\cs Proof of Theorem 2.}
A consequence of Lemma 1 is that for $\l\ge J+\overline C_\e$
the set $\D_V$ is either empty or the whole $\L$, \ie, there are no contours
in the ground state. This, together with the discussion
in the previous section implies the result of Theorem 2 in the case
$\l\ge J+\overline C_\e$. Moreover, a consequence of Lemma 2 is that
for $J>\overline C_\e$ and $\l\ge 0$ the ground state has no
FM-contours, \ie, it is ferromagnetic. Since $\overline C_\e\ge J_0$ this
in particular means that under the same conditions the conclusions of Theorem 2
are valid (see Remark 1 after Theorem 2). Theorem 2 follows by the combination
of these two results, choosing $C_\e>3\overline
C_\e$. \\
\\
{\cs Proof of Lemma 1 and 2}.
Let $\g=\G(\d)$ and note that
$H(\ul{\vec S}_\L)-H(\ul{\vec T}_\L)$ is given bounded below by $(\l-J)|\g|$
plus the difference
between the self-energies of $\ul{\vec S}_\d$ and $\ul{\vec T}_\d$ plus the
difference between the inside-outside interactions of the spins in $\d$ with
the spins in $\d^c$. By the
exponential decay of the potential and the fact that $\ul{\vec T}_\L$
coincides on $\d$ with the $\l=+\io$ ground state,
the first difference is bounded below by a positive constant minus, possibly,
a term of size $\const.|\g|$; similarly the second difference is bounded
above and below by $\const.|\g|$. A computation shows that
these two constants can be bounded above by $c_1\, \log(1/\e)$,
for small $\e$, and by $c_2\e e^{-\e}$, for large $\e$, where $c_1,c_2$
are two suitable constants. So Lemma 1 is proven. The proof of Lemma 2
goes exactly along the same lines: the energy of a state with a non trivial
set of FM-contours, compared to the energy of the FM state, is given by
$\sum_{\g\in\G}J|\g|$ plus the inside-outside energy associated to any
FM-droplet. The latter is bounded below by $-\const.\sum_{\g\in\G}|\g|$,
with the constant bounded as discussed above.

%%%%%%%%%%%%%%%%%%%%%%%%%%%%%%%%%%%%%%%%%%%%%%%%%%%%%%%%%%%%%%%%%%%%%%%%%%
%%%%%%%%%%%%%%%%%%%%%%%%%%%%%%%%%%%%%%%%%%%%%%%%%%%%%%%%%%%%%%%%%%%%%%%%%%
\section{The Ferromagnetic Case: Finite $\l$ and Zero Mass.}
\setcounter{equation}{0}\label{sec6}
%%%%%%%%%%%%%%%%%%%%%%%%%%%%%%%%%%%%%%%%%%%%%%%%%%%%%%%%%%%%%%%%%%%%%%%%%%
%%%%%%%%%%%%%%%%%%%%%%%%%%%%%%%%%%%%%%%%%%%%%%%%%%%%%%%%%%%%%%%%%%%%%%%%%%

In this section we want to discuss how to generalize the Peierls' estimate
of Lemma 1 to the case $\e=0$.
In this section we shall only consider the case $0\le J<J_0$, with $J_0$
defined as in (\ref{2.4}). The main result of this
section is a generalization of Lemma 1 to the massless case.
Its proof requires the use of the screening properties of the $\l=+\io$
ground state (related to its striped nature, in particular to the fact that
its total polarization is vanishing).
In order to describe the result we also need to introduce the notion of
{\it simple droplet}: we shall say that a droplet $\d$ is simple
if either it is simply connected or it winds around the torus and its
complement is connected. One crucial property of simple droplets $\d$ we shall
need is that the number of sites in $\d$ at a fixed lattice
distance $d$ from $\d^c$ is bounded above by $|\G(\d)|$.
Note also that any collection of
droplets $\DD$ associated to some spin state $\vec{\ul S}_\L$
contains at least one simple droplet.

We are now ready to state the main result of this section.\\
\\
{\bf Lemma 3} (Peierls' estimate - massless case).
{\it Let $\e= 0$. If $0\le J<J_0$ and $\l\ge \const.\, h^*(J)$, for a
suitable constant and with $h^*(J)$ the minimizer of the r.h.s. of
(\ref{2.5}), then the following is
true. Let $\ul{\vec S}_\L$ be a spin configuration in $\L$ and let
$\d\in\DD$ be one of its droplets.
If $\d$ is a simple $V$-droplet (resp. $H$-droplet), then
the spin configuration $\vec{\ul T}_\L$ coinciding with $\ul{\vec S}_\L$ on
$\d^c$ and with $\vec{\ul S}^H$ (resp. $\vec{\ul S}^V$) on $\d$ satisfies
$H(\ul{\vec S}_\L)-H(\ul{\vec T}_\L)> 0$.}\\

{\cs Proof of Theorem 1.}
An immediate consequence of Lemma 3 is that in the ground state $|\DD|=1$.
In fact if by contradiction
the ground state configuration $\ul{\vec S}_\L$ had $|\DD|>1$, then
it would be possible to reduce the energy by changing $\ul{\vec S}_\L$
into $\ul{\vec T}_\L$ in the way described above (note that we are using that,
as remarked above, any droplet
configuration $\DD$ always contains at least one simple droplet).
Then all spins are either horizontal or vertical and this,
together with the discussion
of Section \ref{sec4}, implies the result stated in Theorem 1.\\

{\cs Proof of Lemma 3.}
With no loss of generality we assume that $\d$ is a $V$-droplet.
We rewrite $\ul{\vec S}_\L$ in the form: $\ul{\vec S}_\L=
\vec{\ul S}_\d\cup\ul{\vec S}_{\d^c}$ where the spins
in $\vec{\ul S}_\d$ are all vertical and in particular they consitute
a maximally connected component of vertical spins.
We rewrite
\be H(\ul{\vec S}_\L)-H(\ul{\vec T}_\L)=H_0(\vec{\ul S}_\d)-H_0(
\vec{\ul S}^H_\d)+H_1(\vec{\ul S}_\d|\vec{\ul S}_{\d^c})-H_1(
\vec{\ul S}^H_\d|\vec{\ul S}_{\d^c})\label{4.0}\ee
where $H_0(\vec{\ul S}_\d)$ is the internal energy of the spins in $\d$
and $H_1(\vec{\ul S}^H_\d|\vec{\ul S}_{\d^c})$ is the inside-outside
interaction between the spins in $\d$ and those in $\d^c$.

Let us consider the auxiliary configuration $\vec{\ul D}_\L=\vec{\ul S}_\d
\cup{\vec{\ul S}}_{\d^c}^V$ coinciding with $\vec{\ul S}_\L$ inside $\d$ and
coinciding with the ($\l=+\io$) vertical ground state $\vec{\ul S}_\L^V$
outside $\d$. (\ref{4.0}) can be rewritten as
\bea H(\ul{\vec S}_\L)-H(\ul{\vec T}_\L)&=&\big[
H(\vec{\ul D}_\L)-H(\vec{\ul S}^V_\L)\big]+
H_1(\vec{\ul S}_\d|\vec{\ul S}_{\d^c})+\label{4.00}\\
&+&\big[H_0(\vec{\ul S}^V_\d)
-H_0(\vec{\ul S}^H_\d)-H_1(\vec{\ul S}^H_\d|\vec{\ul S}_{\d^c})
+H_1(\vec{\ul S}_\d|\vec{\ul S}^V_{\d^c})
-H_1(\vec{\ul S}^V_\d|\vec{\ul S}^V_{\d^c})
\big]\nn\eea
We shall now separetely estimate the contributions from the three square
brackets in the r.h.s.

By (\ref{3.12}) and defining $\g\=\G(\d)$,
the first term can be bounded from below by
\be H(\ul{\vec D}_\L)-H(\vec{\ul S}_\L^V)\ge -2J|\g|+
\sum_{B: h_B\neq h^*}
\k e^{-\a h^*} h_B+\sum_{<B_1,B_2>:\atop h_{B_1}\neq h_{B_2}=h^*}\k e^{-\a h^*}
(h_{B_1}+h_{B_2})
\label{4.2}\ee
where the notation is the same as in (\ref{3.12}) and we used that
$c_J,d_J\ge \k e^{-\a h^*}$, see lines following (\ref{3.12}).
The second term in (\ref{4.00}) can be bounded below as
\be H_1(\vec{\ul S}_\d|\vec{\ul S}_{\d^c})\ge \l|\g|-
\sum_{B: h_B\neq h^*}|H_{dip}(B_\d|\vec{\ul S}_{\d^c})|
-\sum_{<B_1,B_2>:\atop h_{B_2}=h^*}|
H_{dip}(\{B_1,B_2\}_\d|\vec{\ul S}_{\d^c})|
\label{4.3}\ee
where $H_{dip}(B_\d|\vec{\ul S}_{\d^c})$ is the dipole-dipole interaction
energy between the spins in $B\cap\d$ and $\vec{\ul S}_{\d^c}$,
and similarly $H_{dip}(\{B_1,B_2\}_\d|\vec{\ul S}_{\d^c})$ is
the interaction energy between the spins in $\{B_1\cup B_2\}\cap\d$
and $\vec{\ul S}_{\d^c}$. Note that now in the second sum also pairs
of blocks with both blocks of optimal size are included. The r.h.s. of
(\ref{4.3}) can be bounded below by:
\be \l|\g|-\const.\,\Big[\sum_{\xx\in B: h_B\neq h^*}\frac1{z_\xx}
+\sum_{\xx\in B_1\cup B_2:\atop h_{B_1}\neq h_{B_2}=h^*}\frac1{z_\xx}\Big]
-\sum_{<B_1,B_2>:\atop h_{B_1}=h_{B_2}=h^*}|
H_{dip}(\{B_1,B_2\}_\d|\vec{\ul S}_{\d^c})|\label{4.3a}\ee
where $z_\xx\ge 1$ is the lattice distance between $\xx$ and $\d^c$.
In (\ref{4.3a})
we are only left with the summation over pairs of optimal blocks.
Some of these pairs can be at a distance from $\d^c$ smaller
than $h^*(J)$, and the contribution from all such pairs is bounded below by
$-\const.\,h^*(J)|\g|$.
Now note that a pair of nearest neighbor optimal blocks
has vanishing polarization, so we can bound the contribution of
any optimal pair $\{B_1,B_2\}$ at a distance from $\d^c$ larger than $h^*(J)$
from below by
$$-\const.\,h^*(J)\sum_{\xx\in B_1\cup B_2:\atop h_{B_1}=h_{B_2}=h^*}
\frac1{z_\xx^2}$$
(note that the $\frac1{z_\xx}$ has been now replaced by $\frac1{z_\xx^2}$!).
The total contribution due to pairs of optimal blocks can then be bounded below
as:
$$-\const.\,h^*(J)|\g|-\const.\,h^*(J)
\sum_{d\ge 1}\sum_{\xx\in\d:\atop z_\xx=d}
\frac1{d^2}\ge -\const.\,h^*(J)|\g|$$
where the last constant is in general different from those at l.h.s. and we
used that the number of sites at a distance $d$ from $\d^c$ is bounded
above by $|\g|$ (because $\d$ is a simple droplet). Using this result
in (\ref{4.3a}) we find that, for a suitable $C>0$:
\be H_1(\vec{\ul S}_\d|\vec{\ul S}_{\d^c})\ge \big(\l-C h^*(J)\big)
|\g|-C\,\Big[\sum_{\xx\in B: h_B\neq h^*}\frac1{z_\xx}
+\sum_{\xx\in B_1\cup B_2:\atop h_{B_1}\neq h_{B_2}=h^*}\frac1{z_\xx}\Big].
\ee
Similarly, the contribution from the last square bracket in (\ref{4.00})
is bounded below by $-C h^*(J)|\g|$. The conclusion is that
\be H(\ul{\vec S}_\L)-H(\ul{\vec T}_\L)\ge (\l-C_1 h^*)|\g|
+\sum_{\xx\in B: h_B\neq h^*}\Big(\k e^{-\a h^*}-\frac{C_2}{z_\xx}\Big)
+\sum_{\xx\in B_1\cup B_2:\atop h_{B_1}\neq h_{B_2}=h^*}
\Big(\k e^{-\a h^*}-\frac{C_2}{z_\xx}\Big)\ee
for suitable constants $C_1,C_2>0$. To the purpose of a bound from below we
can trow away all terms with $z_\xx\ge C_2 e^{\a h^*}/\k$ in the two sums.
We are then left with:
\be H(\ul{\vec S}_\L)-H(\ul{\vec T}_\L)\ge (\l-C_1 h^*)|\g|
-3C_2\sum_{\xx\in\d:\atop \k z_\xx\le C_2 e^{\a h^*}}\frac1{z_\xx}.\ee
Since $\d$ is simple, the number of points at a distance $d$ from $\d^c$
is at most $|\g|$. Then the last summation
is bounded below by $-\const.\,|\g|\log(C_2 e^{\a h^*}/\k)$
and the proof of the lemma is concluded.

%%%%%%%%%%%%%%%%%%%%%%%%%%%%%%%%%%%%%%%%%%%%%%%%%%%%%%%%%%%%%%%%%%%%%%%%%%
%%%%%%%%%%%%%%%%%%%%%%%%%%%%%%%%%%%%%%%%%%%%%%%%%%%%%%%%%%%%%%%%%%%%%%%%%%
\section{The Antiferromagnetic Case.}
\setcounter{equation}{0}\label{secanti}
%%%%%%%%%%%%%%%%%%%%%%%%%%%%%%%%%%%%%%%%%%%%%%%%%%%%%%%%%%%%%%%%%%%%%%%%%%
%%%%%%%%%%%%%%%%%%%%%%%%%%%%%%%%%%%%%%%%%%%%%%%%%%%%%%%%%%%%%%%%%%%%%%%%%%

In this section we conclude the proof of Theorem 3. We already proved it in the
$\l=+\io$ case, see Section \ref{sec4B}. So, we now simply need to show that
if either {\it(i)} $|J|$ is larger than some absolute constant $\k_1$ and
$\l=0$ or {\it(ii)} $|J|\le\k_1$ and $\l$ larger than some absolute constant
$\k_2$, then the system always prefers to have the spins all
oriented either out-of-plane or in-plane horizontally or vertically.
If this is the case then Theorem 3 follows, with $\l_0=\k_1+\k_2$.

The easiest case to handle, that we shall treat first, is the case of $|J|$
large and $\l=0$. In this case, both the in-plane and the out-of-plane
minimal energy states display staggered antiferromagnetic order. In analogy
with the FM-contours defined in Section \ref{sec5}, we can introduce the notion
of AF-contours, obtained as union of dual bonds separating nearest neighbor
spins which are either idential or orthogonal. By construction,
the AF-contours separate maximally connected regions of spins displaying
staggered antiferromagnetic order (AF-droplets). Let $\DD_i$ (resp. $\DD_o$)
be the set of in-plane (resp. out-of-plane) AF-droplets and let
$e_{ip}=e_\io^{dip}-2|J|$ and $e_{op}=e_0-2|J|$ be
the in-plane and out-of-plane minimal energies (we recall that $e_\io^{dip}$
and $e_0$ were defined in Section \ref{sec4B}). As discussed in Section
\ref{sec4B} we have $e_{ip}>e_{op}$.
Given a spin configuration $\vec{\ul S}_\L$
whose set of droplets is $\DD_i\cup\DD_o$, its energy can be bounded
below by
\be\sum_{\d\in\DD_i}\Big[e_{ip}|\d|+\Big(\frac{|J|}2-\const.\Big)|\G(\d)|\Big]+
\sum_{\d\in\DD_o}\Big[e_{op}|\d|+\Big(\frac{|J|}2-\const.\Big)|\G(\d)|\Big]
\label{7.1}\ee
where the two (absolute) constants take into account the dipole interaction
energy of the spins inside the droplet with the spins outside (note that this
inside--outside interaction is simply proportional to the length of the
contour, because of the screening effect associated with the antiferromagnetic
phase). By (\ref{7.1}) we see that if $|J|$ is larger than an absolute
constant $\k_1$ then the unique ground state is the out-of-plane staggered
antiferromagentic state.

Let us now turn to the case $|J|\le\k_1$ and $\l$ is larger than some absolute
constant $\k_2$, to be determined below. Given any spin configuration
$\vec{\ul S}_\L$, let us
define $H$-droplets and $V$-droplets as in Section \ref{sec5} and let us also
introduce the notion of $O$-droplets as the maximal connected regions of
out-of-plane spins (the set of $O$-droplets will be denoted by $\DD_O$).
Since the dipole interaction between an in-plane and an out-of-plane spin is
zero, the energy $H(\vec{\ul S}_\L)$ can be rewritten as:
\be H(\vec{\ul S}_\L)=\Big(\frac\l2\sum_{\d\in\DD_H\cup\DD_V}|\G(\d)|
+H_0(\DD_H\cup\DD_V)\Big)+\Big(\frac\l2\sum_{\d\in\DD_O}|\G(\d)|+
H_0(\DD_O)\Big)\=(I)+(II)\label{7.2}\ee
where $H_0(\DD_H\cup\DD_V)$ (resp. $H_0(\DD_O)$) is the interaction energy
among the in-plane (resp. out-of-plane) spins, corresponding to the Hamiltonian
(\ref{2.1}) with $J\le 0$ and $\l=0$. Using the chessboard estimate, $(II)$
can be easily bounded below by $(\l/2)\sum_{\d\in\DD_O}|\G(\d)|+
\sum_{\d\in \DD_O}(e_0-2|J|)|\d|$. In order to bound $(I)$ we follow the same
strategy of Section \ref{sec6}, but we first need to fill the regions occupied
by the $O$-droplets by an auxiliary
configuration of in-plane spins. To this purpose, we define the auxiliary
spin configuration $\vec{\ul T}_\L$, which coincides with $\vec{\ul S}_\L$ on
$\DD_H\cup\DD_V$, and with $\vec{\ul S}_\L^H$ on $\DD_O$. We have:
\be (I)\ge H(\vec{\ul T}_\L)-\sum_{\d\in\DD_O}\Big[
e^*|\d|+\const. |\G(\d)|\Big]\;,\label{7.3}\ee
where $e^*$ is the specific energy of $\vec{\ul S}_\L^H$ (in terms of the
notation introduced in Section \ref{sec4B}, $e^*=e^{AF}(h^*)$ and $h^*\=
h^*(|J|)$).
Now $H(\vec{\ul T}_\L)$ can be bounded below by the method discussed
in Section \ref{sec6}. However now the role of ``blocks of the wrong
size'' is played by ``elementary defects'',
\ie, sequences of 3 contiguous identical spins. Note in fact
that none of the minimal energy states discussed in Section \ref{sec4B}
contains such subconfiguration of spins and a straightforward computation shows
that the state obtained from an elementary defect by repeated reflections has
specific energy larger than $e^{AF}(h)$, for all $h\ge 1$. This makes
the present discussion much simpler than the discussion of Section \ref{sec6}.
The result is that, as long as $\l$ is larger than an absolute constant, we
have that $H(\vec{\ul T}_\L)\ge |\L|e^*$. Combining this bound
with (\ref{7.2}) and (\ref{7.3}) we finally conclude that
\be H(\vec{\ul S}_\L) \ge (e_0-2|J|)\sum_{\d\in\DD_O}|\d|+e^*
\sum_{\d\in\DD_H\cup\DD_V}|\d|\label{7.4}\ee
and this concludes the proof of (\ref{AFspecific}) and of Theorem 3.

%%%%%%%%%%%%%%%%%%%%%%%%%%%%%%%%%%%%%%%%%%%%%%%%%%%%%%%%%%%%%%%%%%%%%%%%%%
%%%%%%%%%%%%%%%%%%%%%%%%%%%%%%%%%%%%%%%%%%%%%%%%%%%%%%%%%%%%%%%%%%%%%%%%%%
\section{Concluding Remarks.}
\setcounter{equation}{0}\label{sec7}
%%%%%%%%%%%%%%%%%%%%%%%%%%%%%%%%%%%%%%%%%%%%%%%%%%%%%%%%%%%%%%%%%%%%%%%%%%
%%%%%%%%%%%%%%%%%%%%%%%%%%%%%%%%%%%%%%%%%%%%%%%%%%%%%%%%%%%%%%%%%%%%%%%%%%

We rigorously proved existence of periodic striped order with periods of length
$h(J)$ in the ground states of a 2D system of dipoles with restricted
orientations, with 3D dipole-dipole long range interactions competing
with a nearest neighbor {\it ferromagnetic} exchange interaction of strength
$J$. We also considered such system with an {\it antiferromagnetic} exchange
$|J|$, in which case we proved existence
of a reorientation transition from an in-plane to an out-of-plane ordered
ground state, as $|J|$ increased. Finally, we gave an example of
soft striped order in the form of 1D sinusoidal spin wave \cite{Nussinov}.

Unfortunately, even the ground states of more realistic models used
to describe thin films are
still far from being solved exactly. On the basis of variational arguments and
approximations, one expects spontaneous formation of mesoscopic stripes in
anisotropic systems of out-of-plane spins interacting via the dipole-dipole
(or Coulomb) long-range interaction and a short range FM exchange, like those
considered in \cite{SS99,MWRD95,AWMD95,LEFK94,CEKNT96}. It is also expected
that in the presence of a sufficiently strong
uniform magnetic field, oriented perpendicular to the plane, the ground state
should exhibit periodic order in the form of bubbles, \ie, domains of spins
parallel to the external field of quasi--circular shape \cite{GD82}.
Striped or bubble patterns are also expected on the basis of an effective (mean
field) free energy functional in 2D electron gases
\cite{JKS05,SK04,SK06}, Langmuir monolayers and liquid crystals \cite{SWBK93}.

Even less is known rigorously for positive temperatures. In particular
it is unclear, even on a heuristic level, whether the expected striped or
bubbled order for discrete spins should have strict long range
order (LRO), or rather quasi--long range order (QLRO) characterized by
order on short scales and a power law
decay of the order parameter correlation functions.
One of the few rigorous results, which we are aware of, about the
positive temperature behavior of this class of systems is the recent proof
by Biskup, Chayes and Kivelson of the absence of ferromagnetism in
$d$--dimensional Ising models with long range repulsive interactions, decaying
as $1/r^p$, $d<p\le d+1$, interactions \cite{BCK07}. 

The models discussed here are related to a class of systems with Kac
potentials (\ie, long range potentials of the form $\gamma^d v(\gamma
r)$) considered by Lebowitz and Penrose in \cite{LP66}. There they
computed the free energy density of a system of particles interacting
via a short range interaction, favoring phase segregation, and a long
range nonnegative definite Kac potential.  They showed that in the
limit $\g\to 0$ there is no phase transition in the thermodynamic
sense, even though the pair distribution function has the form
characterizing a phase transition, at least over length scales much
smaller than $\g^{-1}$. They concluded that the repulsive Kac
potential causes the distinct phases of a normal first-order phase
transition to break into droplets, or froth, of characteristic length
large compared to the range of the short range potential and small
compared to $\g^{-1}$.  We do not know the scale of these domains, or
even whether they destroy the first-order transition in 2D and 3D when 
$\gamma$ is small but finite.

In 1D we can show, using the method of reference \cite{GLL06}, for
$H=-J \sum \sigma_i\sigma_{i+1} + \sum_{i,j}\gamma \exp\{-\gamma
|i-j|\}$, that for $\gamma \to 0$ the ground states are periodic with
period proportional to $\gamma^{-1} (J\gamma)^{1/3}$, \ie, on the
macroscopic scale $\gamma^{-1}$ the period goes as $(J\gamma)^{1/3}$.
This is very reminiscent of what happens to the minimizers of
continuum energy functionals used to model microphase separation of diblock 
copolymers and many other physical systems 
\cite{ohtakawasaki,muller,choksi,albertimuller,chenoshita,cgll}.

For Heisenberg spins with long
range interactions the Hohenberg-Mermin-Wagner argument is not applicable and
it could very well be that the long range tails stabilize phases against
thermal fluctuations, even in two dimensions.
This issue has been discussed in some detail for
the case of rotators in two dimensions, interacting via a 3D pure dipole-dipole
interaction \cite{AWDM01,CRRT00}: this is a case where
linear spin-wave theory predicts non-existence of LRO, while
non-linear corrections and renormalized spin-wave theory seem to suggest that
LRO survives at positive temperatures.

For scalar fields describing the local magnetization or electron density
in an effective free energy functional theory, the general expectation is that
in the presence of an anisotropy term
(possibly induced by the underlying crystalline structure) QLRO should survive
at positive temperatures. On the contrary
even QLRO should generally be destroyed by thermal fluctuations in the case of
isotropic interactions \cite{TKNV05}. The issue is rather subtle, however:
analyses based on a Hartree approximation would generically
predict the presence of a first order phase transition from a high temperature
disordered phase to a low temperature striped phase \cite{B75}.
If both predictions are correct, this class of models would exhibit a
rather peculiar first order phase transition from a disordered state to a
locally ordered state without strict LRO.
The issue has been investigated
by Jamei, Kivelson and Spivak \cite{SK04,JKS05,SK06}
in the context of 2D electron gases and by Tarjus, Kivelson, Nussinov and
Viot \cite{TKNV05} in the context of
the frustration-based appoach to the glass transition in supercooled
liquids and structural glasses.
Jamei, Kivelson and Spivak argued that for 2D electron gases, in the presence
of Coulomb interactions, first order transitions (as those predicted by
Brazovskii) are not allowed and a sequence of transition between
different mesoscopic patterned states should generically appear instead.
On the contrary, as discussed in \cite{TKNV05}, numerical simulations of
frustrated spin models with Coulomb interaction predict a finite
jump in the energy at a critical line separating the paramagnetic state from
a locally order striped state.

Refinements of the mean field or variational arguments seem very difficult:
the natural effective
continuum theories describing systems whose interaction has a Fourier transform
with a non-trivial minimum in $k$-space are, at least naively,
non-renormalizable \cite{HS95,S06}. It would be
very interesting to provide convincing arguments for the
patterned states to be (globally) stable against the presence of thermal or
quantum fluctuations, as well as against ``generic'' perturbations in the form
of long or short range interactions.

\acknowledgments

We thank L. Chayes, S. Kivelson, G. Tarjus and P. Viot for useful discussions
and comments and we thank Z. Nussinov for informing us  that the basic results
about $o(n)$ models used in section \ref{sec3} are contained in his thesis 
\cite{Nussinov}.
We also thank A. De Masi, O. Penrose and E. Presutti for helpful discussions 
about the Kac potential.
The work of JLL was partially supported by NSF Grant DMR-044-2066 and by AFOSR
Grant AF-FA 9550-04-4-22910. The work of AG and EHL
was partially supported by U.S. National Science Foundation
grant PHY-0652854.

%%%%%%%%%%%%%%%%%%%%%%%%%%%%%%%%%%%%%%%%%%%%%%%%%%%%%%%%%%%%%%%%%%%%%%%%%%%%
\appendix
\section{A new proof of periodic order in reflection-positive 1D Ising
systems}\lb{A}
\setcounter{equation}{0}
\renewcommand{\theequation}{\ref{A}.\arabic{equation}}
%%%%%%%%%%%%%%%%%%%%%%%%%%%%%%%%%%%%%%%%%%%%%%%%%%%%%%%%%%%%%%%%%%%%%%%%%%%%
%%%%%%%%%%%%%%%%%%%%%%%%%%%%%%%%%%%%%%%%%%%%%%%%%%%%%%%%%%%%%%%%%%%%%%%%%%%%

We mentioned in the Introduction that the proofs of Theorems 1 and 2 in
\cite{GLL06} contain a technical error, which was caused by overlooking some
exponentially small terms that came from the choice of periodic boundary
conditions. In this Appendix we show how to repair this error. We exploit a
new method, which is simpler than the one used in Section III of \cite{GLL06}
and which does not make use of periodic, but rather of open boundary
conditions, as will be discussed below. The proof given in this Appendix also
implies the results discussed in Section \ref{sec4}.

First of all, let us point out the mistake in \cite{GLL06}: a term
\be\sum_{1\le i<j\le M/2}(-1)^{i-j}(1-e^{-\a h_i})(1-e^{-\a h_j})
e^{-2\a N}\prod_{i\le k\le j} e^{\a h_k}\label{A.00}\ee
is missing in the definition of $H_R(\a,\ul{h}_R)$ in (I.9),
\ie, in Eq.(9) of \cite{GLL06}. Taking into account this term,
one can check that, whenever the total length $N_L=
\sum_{\frac{M}2<i\le 0} h_i$ of the blocks on the left is different from
the total length $N_R=\sum_{1\le i\le \frac{M}2} h_i$ of the blocks on the
right, the new energy functions $H(\a,(\ul{h}_L,\hat\th\ul{h}_L))$ and
$H(\a,(\hat \th\ul{h}_R,\ul{h}_R))$ obtained after the first reflection,
see Eq.(11) of \cite{GLL06}, are not periodic. For this reason it does not seem
possible to repeatedly reflect, as one must do in order to obtain the
checkerboard estimate, see \cite{GLL06}, Eq.(12).\\

Let us now show how to correct the proofs of Theorems 1 and 2 in \cite{GLL06}.
Let us denote Eq.(x) of \cite{GLL06} by (I.x) and let $H_N$ be the periodic
Hamiltonian in (I.1):
\begin{equation}H_N(\ss)=-J\sum_{i=-N+1}^N\s_i\s_{i+1}
+\sum_{-N+1\le i<j\le N}
\s_i\, J_p(j-i)\; \s_j\;,\qquad J_p(j-i)=\sum_{n\in \zzz^d}
\fra1{|i-j+2n N|^p}\label{1.1}\end{equation}
where $\s_{N+1}\=\s_{-N+1}$ and $p>1$. As discussed in \cite{GLL06},
Theorems 1 and 2 are consequences of the {\it chessboard estimate}, used in
\cite{GLL06} and stating that
\be H(h_1,\ldots,h_n)\ge \sum_{i=1}^n h_i e(h_i)\;, \qquad \sum_{i=1}^nh_i=N
\;,\label{A.9}\ee
where $H(h_1,\ldots,h_n)$ is the energy $H_N(\ss)$ of a spin configuration
whose corresponding block configuration is $\ul{h}=\{h_1,\ldots,h_n\}$ and
$e(h)$ is the energy per site of the infinite system with a periodic 
configuration of blocks all of
the same size and alternating sign (let us recall that a block is a maximal
sequence of spins all of the same sign). In the following we shall also need to
introduce the analogue of $H(\ul{h})$ with {\it open boundary
conditions}, to be denoted by $H^0(\ul{h})$.
Since the long range potential is summable, we have
\be H(\ul{h})=\lim_{m\to\io}\frac1m\,
H^0(\overbrace{\ul{h},\ldots,\ul{h}}^{m {\rm times}}) \ee
and we find that the chessboard estimate (\ref{A.9}) is a consequence of the
following:\\

{\bf Chessboard estimate with open boundary conditions.}
{\it Given a finite sequence of blocks $\uA_i=\{h_1^i,\ldots,h_{m_i}^i\}$,
let us denote by $\th\uA_i=\{h_{m_i}^i,\ldots,h_1^i\}$ the reflection of
$\uA_i$ (with the sign of the spins in the reflected blocks
being opposite to what they were originally).  By $e(\uA_i)$ we denote
the infinite volume energy per site of the
configuration $(\ldots,\uA_i,\th\uA_i,
\uA_i,\th\uA_i,\ldots)$ and let $a_i=\sum_{j=1}^{m_i}h^i_j<\io$.
Then, for any collection $\uA_0,\ldots,\uA_{n+1}$, with
$n\ge 1$ and $\sum_{i=0}^{n+1}a_i=N$, we have
\be H^0(\uA_0,\uA_1,\ldots,\uA_n,\uA_{n+1})\ge
(a_0+a_{n+1}) e_0 +\sum_{i=1}^n a_n e(\uA_n)\;,\label{A.10}\ee
where $e_0$ is the infinite volume specific ground state energy.}\\

{\cs Proof.} Let us first note that, for any sequence of blocks $\uA$,
$e(\uA)\ge a e_0$, where $a=\sum_{h\in\uA}h$. This can be proven as follows.
Denote by $\ss_a$ the spin configuration corresponding to $\uA$, so that
$H^0(\uA)=H^0_a(\ss_a)$. If $\o=\pm$, we have
$H^0_{2a}(\ss_a,\o\ss_a)=2H^0_a(\ss_a)+\o E_{int}(\ss_a;\ss_a)$, where
$(\ss_a,\o\ss_a)$ is the spin configuration of length $2a$ obtained by
juxtaposing $\ss_a$ and $\o\ss_a$, and $\o E_{int}(\ss_a;\ss_a)$ is the
interaction energy between the two halves. For one of the two choices $\o=\pm$
this interaction energy is nonpositive, so that $2H^0_a(\ss_a)\ge
\min_{\ss} H^0_{2a}(\ss)$. Iterating, we find:
\be H^0_a(\ss_a)\ge \lim_{m\to\io}2^{-m}\min_{\ss}
H^0_{2^m a}(\ss)=a e_0\;,\label{A.10a}\ee
which is the desired estimate.

Next, let us recall that reflection
positivity of the long range potential $1/r^p$ implies the following basic
estimate.
\\

{\bf Lemma A.1.} {\it Given two finite sequences of blocks $\uA_-=\{h_{-M+1},
\ldots,h_0\}$ and $\uA_+=\{h_1,\ldots,h_N\}$, with $M,N\ge 1$,
let $\th\uA_-=\{h_0,\ldots, h_{-M+1}\}$ and $\th\uA_+=\{h_N,\ldots,h_1\}$
be their reflections. Then we have:
\be H^0(\uA_-,\uA_+)\ge
\frac12 H^0(\th\uA_+,\uA_+)+\frac12H^0(\uA_-,\th\uA_-)\;.\label{A.1a}\ee}

Now we are ready to prove (\ref{A.10}). We proceed by induction.
If $n=1$ in (\ref{A.10}), then by reflection positivity, \ie, by Lemma A.1,
we have:
\be H^0(\uA_0,\uA_1,\uA_2)\ge \frac12 H^0(\th\uA_2,\uA_2) +\frac12
H^0(\uA_0,\uA_1,\th\uA_1,\th\uA_0)\label{A.11}\ee
By (\ref{A.10a}), the first term in the r.h.s. can be bounded from below by
$a_2 e_0$.
The second term can be bounded by a second reflection:
\be \frac12 H^0(\uA_0,\uA_1,\th\uA_1,\th\uA_0)\ge \frac14 H^0(\th\uA_0,\uA_0) +
\frac14 H^0(\uA_0,(\uA_1)^{\otimes 4},\th\uA_0)\label{A.13}\ee
where by definition $(\uA_1)^{\otimes 4}=(\uA_1,\th\uA_1,\uA_1,\th\uA_1)$.
By (\ref{A.10a}),
the first term in the r.h.s. of (\ref{A.13})
can be bounded from below by $a_0 e_0/2$, so we end up with:
\be H^0(\uA_0,\uA_1,\uA_2)\ge a_2 e_0+\frac12 a_0 e_0 +
\frac14 H^0(\uA_0,(\uA_1)^{\otimes 4},\th\uA_0)\label{A.14}\ee
Iterating we find:
\be H^0(\uA_0,\uA_1,\uA_2)\ge a_2 e_0 + a_0 e_0 \sum_{n\ge 1}2^{-n}
+\lim_{n\to\io} 2^{-n} H^0(\uA_0,(\uA_1)^{\otimes 2^n},\th\uA_0)\label{A.15}\ee
Note that the last term is equal to $a_1 e(\uA_1)$, so the desired
bound is proven for $n=1$.

Let us now assume by induction that the bound is valid for all $1\le k\le n-1$,
$n\ge 2$, and let us prove it for $k=n$. There are two cases.

\\(a) $n=2p$ for some $p\ge 1$. If we reflect once we get:
\bea && H^0(\uA_0,\uA_1,\ldots,\uA_{2p},\uA_{2p+1})\ge\label{A.16} \\
&& \quad \ge \frac12
H^0(\th\uA_{2p+1},\ldots,\th\uA_{p+1},\uA_{p+1},\ldots,\uA_{2p+1})+ \frac12
H^0(\uA_0,\uA_1,\ldots,\uA_{p},\th\uA_p,\ldots,\th\uA_1,\th\uA_0) \nn\eea
If we now regard $\uA_{p+1}'\=(\th\uA_{p+1},\uA_{p+1})$ and
$\uA_p'\=(\uA_{p},\th\uA_p)$ as two new sequences of blocks,
the two terms in the r.h.s. of (\ref{A.16}) can be regarded as two terms with
$n=2p-1$ and, by the induction assumption, they satisfy the bounds:
\bea && H^0(\th\uA_{2p+1},\ldots,\uA_{p+1}',\ldots,\uA_{2p+1})
\ge 2 a_{2p+1} e_0 + 2 \sum_{i=p+1}^{2p} a_i e (\uA_i) \\
&&H^0(\uA_0,\uA_1,\ldots,\uA_{p}',\ldots,\th\uA_1,\th\uA_0) \ge
2 a_{0} e_0 + 2 \sum_{i=1}^{p} a_i e (\uA_i) \label{A.17}\eea
where we used that, by construction, $a'_{p}=2 a_{p}$, $a'_{p+1}=2a_{p+1}$,
$e(\uA'_{p})=e(\uA_{p})$ and $e(\uA'_{p+1})=e(\uA_{p+1})$.
Therefore, the desired bound is proven.

\\(b) $n=2p+1$ for some $p\ge 1$. If we reflect once we get:
\bea && H^0(\uA_0,\uA_1,\ldots,\uA_{2p+1},\uA_{2p+2})\ge\label{A.18} \\
&& \frac12
H^0(\th\uA_{2p+2},\ldots,\th\uA_{p+2},\uA_{p+2},\ldots,\uA_{2p+2})+ \frac12
H^0(\uA_0,\uA_1,\ldots,\uA_{p+1},\th\uA_{p+1},\ldots,\th\uA_1,\th\uA_0) \nn\eea
The first term in the r.h.s. corresponds to $n=2p$ so by the induction
hypothesis it is bounded below by $a_{2p+2} e_0 +\sum_{i=p+2}^{2p+1} a_1
e(\uA_i)$. As regards the second term, using reflection positivity again,
we can bound it from below by
\be \frac14
H^0(\uA_0,\uA_1,\ldots,\uA_{p},\th\uA_{p},\ldots,\th\uA_1,\th\uA_0)+\frac14
H^0(\uA_0,\uA_1,\ldots,\uA_{p},(\uA_{p+1})^{\otimes 4},\th\uA_{p},
\ldots,\th\uA_1,\th\uA_0)\label{A.19}\ee
By the induction hypothesis, the first term is bounded below by
$a_0 e_0/2 +(1/2)\sum_{i=1}^p a_i e(\uA_i)$, and the second can be bounded
using reflection positivity again. Iterating we find:
\bea && H^0(\uA_0,\uA_1,\ldots,\uA_{2p+1},\uA_{2p+2})\ge\label{A.20} \\
&& \quad \ge a_{2p+2} e_0 +\sum_{i=p+2}^{2p+1} a_i e(\uA_i)
+ \Big(\sum_{n\ge 1}2^{-n}\Big) \Big(a_0 e_0+\sum_{i=1}^p a_i e(\uA_i)\Big)
+\nn\\
&&\hspace{2.truecm} +\lim_{n\to\io} 2^{-n}
H^0(\uA_0,\uA_1,\ldots,\uA_{p},(\uA_{p+1})^{\otimes 2^n},\th\uA_{p},
\ldots,\th\uA_1,\th\uA_0)\;. \nn\eea
Note that the last term is equal to $a_{p+1} e(\uA_{p+1})$, so (\ref{A.20})
is the desired bound. This concludes the proof of the chessboard estimate
with open boundary conditions and, as mentioned above, of (\ref{A.9})
and of Theorems 1 and 2 in \cite{GLL06}. \qed

%\\
%\\
%{\bf References.}\\
%[DMW] K. De'Bell, A. B. MacIsaac and J. P. Whitehead: {\it
%Dipolar effects in magnetic thin films and quasi-two-dimensional systems},
%Rev. Mod. Phys. 72, 225 - 257 (2000).
%\\
%[FILS1] J. Fr\"ohlich, R. B. Israel, E. H. Lieb and B. Simon:
%{\it Phase Transitions and Reflection Positivity. I. General Theory and
%Long Range Lattice Models}, Commun. Math. Phys. 62, 1-34 (1978).
%\\
%[FILS2] J. Fr\"ohlich, R. B. Israel, E. H. Lieb and B. Simon:
%{\it Phase Transitions and Reflection Positivity. II. Lattice Systems with
%Short-Range and Coulomb Interactions}, J. Stat. Phys. 22, 297-347 (1980).
%\\
%[FS] J. Fr\"ohlich and T. Spencer: {\it On the statistical mechanics of
%classical Coulomb and dipole gases}, Journal of Statistical Physics 24,617-701
%\\
%[GLL] A. Giuliani, J. L. Lebowitz and E. H. Lieb: {\it Ising models with
%long-range antiferromagnetic and short-range ferromagnetic interactions},
%Phys. Rev. B {\bf 74}, 064420 (2006).
%\\
%[GR] I. S. Gradshteyn, I. M. Ryzhik: {\it
%Table of Integrals, Series, and Products}, Sixth edition, Academic Press
%(2000).

\end{document}